\documentclass[sigconf]{acmart}

\AtBeginDocument{%
  \providecommand\BibTeX{{%
    \normalfont B\kern-0.5em{\scshape i\kern-0.25em b}\kern-0.8em\TeX}}}

\copyrightyear{2023} 
\acmYear{2023} 
\setcopyright{acmlicensed}
\acmConference[CHI '23]{Proceedings of the 2023 CHI Conference on Human Factors in Computing Systems}{April 23--28, 2023}{Hamburg, Germany} 
\acmBooktitle{Proceedings of the 2023 CHI Conference on Human Factors in Computing Systems (CHI '23), April 23--28, 2023, Hamburg, Germany} 
\acmPrice{15.00} 
\acmDOI{10.1145/3544548.3581564} 
\acmISBN{978-1-4503-9421-5/23/04} 

\usepackage{quoting}
\begin{document}

\title[Autonomous Vehicle Safety Drivers' Lived Experiences]{Work with AI and Work for AI: Autonomous Vehicle Safety Drivers' Lived Experiences}

\author{Mengdi Chu}
\orcid{0000-0003-0533-7801}
\affiliation{%
  \institution{Institute for AI Industry Research, Tsinghua University}
  \city{Beijing}
  \country{China}
}
\email{mengdichu@outlook.com}

\author{Keyu Zong}
\orcid{0000-0003-0927-6129}
\affiliation{%
  \institution{Institute for AI Industry Research, Tsinghua University}
  \city{Beijing}
  \country{China}
}
\email{zongky@mails.ccnu.edu.cn}

\author{Xin Shu}
\orcid{0000-0002-8898-9698}
\affiliation{%
  \institution{Institute for AI Industry Research, Tsinghua University}
  \city{Beijing}
  \country{China}
}
\email{mr_comfort@163.com}

\author{Jiangtao Gong}
\authornote{Corresponding author}
\orcid{0000-0002-4310-1894}
\affiliation{%
  \institution{Institute for AI Industry Research, Tsinghua University}
  \city{Beijing}
  \country{China}
}
\email{gongjiangtao2@gmail.com}

\author{Zhicong Lu}
\orcid{0000-0002-7761-6351}
\affiliation{%
  \institution{Department of Computer Science, City University of Hong Kong}
  \city{Hong Kong}
  \country{China}
}
\email{zhiconlu@cityu.edu.hk}

\author{Kaimin Guo}
\orcid{0000-0002-7381-8925}
\affiliation{%
  \institution{Institute for AI Industry Research, Tsinghua University}
  \city{Beijing}
  \country{China}
}
\email{15536606362@163.com}

\author{Xinyi Dai}
\orcid{0000-0001-7718-5259}
\affiliation{%
  \institution{Institute for AI Industry Research, Tsinghua University}
  \city{Beijing}
  \country{China}
}
\email{daixinyi53@foxmail.com}

\author{Guyue Zhou}
\orcid{0000-0002-3894-9858}
\affiliation{%
  \institution{Institute for AI Industry Research, Tsinghua University}
  \city{Beijing}
  \country{China}
}
\email{zhouguyue@air.tsinghua.edu.cn}

\renewcommand{\shortauthors}{Chu, et al.}

\begin{abstract}
The development of Autonomous Vehicle (AV) has created a novel job, the safety driver, recruited from experienced drivers to supervise and operate AV in numerous driving missions. Safety drivers usually work with non-perfect AV in high-risk real-world traffic environments for road testing tasks. However, this group of workers is under-explored in the HCI community. To fill this gap, we conducted semi-structured interviews with 26 safety drivers.
Our results present how safety drivers cope with defective algorithms and shape and calibrate their perceptions while working with AV. We found that, as front-line workers, safety drivers are forced to take risks accumulated from the AV industry upstream and are also confronting restricted self-development in working for AV development. We contribute the first empirical evidence of the lived experience of safety drivers, the first passengers in the development of AV, and also the grassroots workers for AV, which can shed light on future human-AI interaction research.
\end{abstract}

\ccsdesc[500]{Human-centered computing~Empirical studies in collaborative and social computing}

\keywords{autonomous driving, human-AI interaction, AI perception, AI labor, responsible AI}

\maketitle

\section{Introduction}

The arrival of autonomous vehicle technologies is expected to revolutionize human daily transportation and promises to enhance road safety, comfort, and mobility.
This rapidly growing field has established a thriving industry in just a matter of a decade~\cite{banks2018keeping,takacs2018assessment}. "Safety Driver" is born of this trend. For technical, legislative, and ethical reasons, fully automated vehicles have not yet been widely implemented, human supervision of AV will still be needed for a long period~\cite{mcgehee2016review,Kalra2016DrivingTS}. Hence, AV companies recruit experienced drivers as safety drivers to supervise and operate autonomous cars to ensure safety and conformity. Safety drivers typically have close and long-term interactions with autonomous vehicles in real-world scenarios. Understanding their practices, experiences, and challenges when working with highly automated systems can offer a glimpse into the upcoming autonomous society and inspire research on human-AI interaction. However, this group is under-explored in the HCI community. Most studies on human-autonomous vehicle interaction are based on laboratory environments or short-term observations that may be isolated from real-world practices and few studies have investigated the day-to-day interactions between drivers and highly automated systems. To fill this gap, we conducted semi-structured interviews with 26 safety drivers.

We explored the following research questions in this study:
\begin{enumerate}
    \item What are the safety drivers' work practices?
    \item How do safety drivers perceive, understand, and partner with AV technologies in working with AI?
    \item What are the experiences and challenges faced by safety drivers in working for AI?
\end{enumerate}
    
We drew a picture of safety drivers' lived experiences and presented how safety drivers perceive, understand, and work with AV technologies:
\begin{itemize}
    \item We examined how individuals with limited knowledge of AV form and adjust their perception of AV, and identified the factors that shape their perception.
    \item We investigated the transition of control between the safety driver and the autonomous system and uncovered tensions between organizational tendencies and individual tendencies.
    \item We found safety drivers' takeover decision-making characteristics in high-risk emergency situations.
    \item We revealed the learning preferences of safety drivers: practices than theories; try and fail than smooth processes; tangible and visible than abstract and invisible; and interaction with colleagues than taught lessons. 
\end{itemize}

We also presented their work experiences, challenges, and well-beings while working for AV industries:
\begin{itemize}
    \item We introduced their distinctive work experiences brought about by AV. 
    \item We dove into the ambiguous responsibility allocation and moral dilemmas in their work. 
    \item We revealed the well-being challenges faced by safety drivers: assuming risks generated by the upstream AV industries, limited opportunities for personal growth, and marginalization.
\end{itemize}

We compared our results with previous studies and explored the potential to enhance human-vehicle partnerships and improve worker experiences. Our research contributes the first empirical evidence of long-term interactions with autonomous vehicles in real-world scenarios. Furthermore, as a real world example of human-AI interactions, this study can serve as an analogy for broader human-AI research that involves real-world implications and provide insights for future human-AI studies.
  
\section{Background}

The Society of Automotive Engineers (SAE) established defined 6 levels of AV to distinguish the responsibilities between the driver and the vehicle, ranging from Level 0 (no automation) to Level 5 (full automation)~\cite{sae2014taxonomy}. Level 1 and Level 2 are commonly referred to as "advanced driver-assistance systems" (ADAS), in which the human driver is in charge of driving tasks and receives assistance from the automation system; Level 3 is able to carry out critical driving tasks under certain conditions, and human drivers are supposed to continuously monitor and take control at all times; Level 4 automation is capable of performing all dynamic driving tasks under specific conditions, and a human driver may take control as necessary; At the highest level, Level 5  is able to perform all driving tasks under any circumstances without any intervention from the driver~\cite{takacs2018assessment,drexler2019handover}. The fully automated system without any human intervention is the desired future of AV. However, this is not attainable in the short term, for technical, legislative, and ethical reasons~\cite{mcgehee2016review,Kalra2016DrivingTS}. Human supervision for automated vehicles is still indispensable.

Generally, autonomous vehicles must undergo thorough public road testing to assess their viability and safety before being made available to the public~\cite{abu2022synthesis}. 
Regions around the world have enacted laws and regulations to guide AV real-road testing. Although regulations vary among countries, most of them require that each autonomous vehicle be equipped with at least one human supervisor, who is responsible for monitoring the vehicle to ensure that the car is driving safely and adhering to traffic laws, and they must be ready to take over the autonomous vehicle if necessary~\cite{wang2020safety}.

Up to the time of this study conducted, under Chinese legal provisions, AV organizations have to ensure the car is under the supervision of human drivers in public road testing, regardless of the level of automation being tested.
Driven by the fast-paced development of AV technology and the desire to trial it in real-world conditions, more and more AV organizations began conducting tests on public roads, which gave birth to a new occupation, "Safety Driver." AV organizations recruit experienced drivers to supervise and operate AV for numerous driving missions to ensure vehicle safety and conformity on public roads. Safety drivers usually get along with AV closely and long-termly in real-world environments. To some extent, they are the first passengers in the development of AV, as well as the future workers in the upcoming automated society.

\section{Related Work}
\subsection{Human-AI Partnership in Highly Automated Vehicles}

Recent advances in AI technologies have led to AI systems becoming more closely embedded in human society, in which people and AI interact in complex ways and work together to collaboratively solve problems and perform specific tasks~\cite{dafoe2021cooperative,ramchurn2021trustworthy}. 
To unlock the potential synergies between humans and machines, a variety of topics in the research and application of human-AI partnerships have been explored, such as shared mental models, goal-alignment, and decision-making between humans and AI systems~\cite{schelble2022let, shively2017human,demir2018team}. 
In highly automated vehicles, the AI system carries out most of the driving tasks that were previously performed by human drivers, however, due to many hurdles to the wide adoption of fully autonomous driving, ranging from reliability to liability issues, human supervision is still essential and the driver needs to partner with the vehicle to finish driving tasks, which changes the relationship between human and AI systems more cooperatively~\cite{tran2018human,xu2023transitioning}. 
Some studies are being performed to explore human-AI collaboration in autonomous vehicle systems, focusing on trust calibration~\cite{wang2020safety,zhang2021human,Dirsehan2020ExaminationOT}, situation awareness~\cite{Lindemann2018CatchMD,Yang2018AnHC,Li2018SwitchedCD}, take-over control~\cite{Kim2018TakeoverPA,Hester2017DriverTO,Deo2020LookingAT} and Human-Machine Interaction (HMI) systems~\cite{Colley2020EffectOV,Ekman2016CreatingAT,Xing2021TowardHC}. 
Shahrdar et al.~\cite{shahrdar2018survey} discussed the misuse of automation caused by distrust and over-trust; Takács et al.~\cite{takacs2018assessment} summarized the challenges in supervising AV: limited human drivers' performance in terms of accuracy, time delay, and complexity, incorrect environment awareness, untimely situation assessment, lack of traffic information, and distraction by non-driving activities; Baltzer et al.~\cite{baltzer2014mediating} presented the conflicts of control authority and responsibility distribution and the challenges of communication and state alignment in the cooperation between humans and automated vehicles. 
Researchers are actively exploring approaches to improving human-vehicle partnerships, including increasing the explainability~\cite{Koo2015WhyDM} and transparency~\cite{Meguia2019PrinciplesOT} of AV systems to enhance the calibration of drivers' trusts~\cite{wang2020safety}, perceptions~\cite{Hewitt2019AssessingPP}, and mental models~\cite{Wiegand2019ID}, and inventing novel HMI to improve drivers' attention and situation awareness~\cite{Du2018VoiceUI,yang2018hmi,llaneras2017strategies}, etc.
Previous studies provided valuable human-vehicle interaction insights, while most of them were built on controlled laboratory environments or short-term observation, which may be isolated from long-term real-world practice. This study aims to fill this gap and investigates the in-the-wild human-AI partnerships in highly autonomous vehicles from the perspective of safety drivers.

\subsection{From Driver Experiences to Autonomy Experiences}

In the field of autonomous driving, understanding the experiences of human drivers - an important part of future traffic by working alongside automation systems - will contribute to future research in human-autonomy interactions and challenges posed by automation~\cite{altendorf2019utility,orii2021perceptions}. 
Previous studies have explored drivers' experiences, perspectives, attitudes, and acceptance of autonomous systems. 
Karvonen et al.~\cite{karvonen2011hidden} investigated drivers' interactions with the automated metro system in Helsinki by conducting observations and interviews. Their findings identified the challenges faced by the drivers, including the demands for dynamic, complex, and uncertain control, the risk of decision-making in exceptional situations, and the monotonous work routine, and also shed light on the importance of considering the human factor in the design and operation of highly automated transportation systems. 
Yang et al.~\cite{yang2018first} presented truck drivers’ on-the-road experience and subjective acceptance of using Cooperative Adaptive Cruise Control (CACC) based on their 160-mile driving experiment in Northern California and identified the factors influencing their acceptance and usage of CACC, such as road environments, traffic conditions, individual differences, etc.
Lee et al.~\cite{lee2017study} conducted fieldwork involving six participants who rode in a prototype autonomous car on real roads for six days to investigate their experience with autonomous vehicles and identify factors that significantly influence passengers' trust in autonomous vehicle including lack of information, unpredictability and value misalignment, etc.
Much of the current empirical research on automated driving experiences has focused on automated mass transit and trucks, and the drivers in these studies typically lack practical experience with day-to-day use of automated systems. There is a lack of empirical research on the actual experience of drivers of highly automated passenger cars, which is seen as one of the most likely autonomous vehicles to be widely used by the general public~\cite{Panagiotopoulos2020AreCR}. 
Moreover, the drivers of autonomous vehicles are also the users of AI systems. Their experiences with AI systems are important to consider in the development and deployment of AI technologies.
Some studies investigated the end-users' subjective perceptions and folk minds of algorithms to gain insights into refining human-AI interactions and improving user understanding and trust of AI~\cite{sonboli2021fairness,long2020ai}. Understanding non-technical users' experiences and perspectives about AI could serve as a valuable heuristic cue to improve the comprehensibility, accessibility, and applicability of AI systems to a much broader group.

\subsection{AI workers and Socio-Technical HCI}

The deployment of emerging technologies, including those related to autonomous vehicles, has significantly impacted the nature of work and the power and social dynamics within workplaces. This shift in technology and work practices requires a rethinking of the relationship between emerging technologies and human workers~\cite{cheon2021human,Baecker2019AutomationWA}. Manyika et al. advocated than compared to discuss whether jobs will be lost, it was important to evaluate how work will change due to the increased interactions of human and autonomous systems~\cite{manyika2017jobs}. Baltrusch et al.~\cite{baltrusch2022human} investigated the impacts of automation technologies on work quality and workers' well-being and identified four factors: cognitive workload, collaboration fluency, trust, and acceptance and satisfaction. Bhoopalam et al.~\cite{kishore2021long} explored the truck drivers' perspectives of autonomous vehicle technologies through conducting focus groups. This study reported the concerns of one of the key stakeholder groups in the transition to AV technology and emphasized the need for careful consideration of the impact on workers and the development of strategies to support them during this transition.
Automation technologies may change human workers from operators to more supervisory roles~\cite{xu2020automation}, promote deskilling for many workers and a need for new skills~\cite{downey2021partial}, create new occupations and opportunities~\cite{wang2020safety}, or increase the marginalization and precariousness of low-skilled workers~\cite{tubaro2020trainer}. 
When developing and deploying new technologies, researchers are supposed to adopt interdisciplinary approaches to form a socio-technical concept that considers not only technical issues, but also ethical and social  implications, which might reveal important design characteristics for the integration of technology into human society~\cite{moniz2014technology,clemmensen2021socio}
Safety driver as the new occupation created by AV industries and seldom documented yet. Learning their work practices could help us foresee how AV technology will be embedded in our society.

\section{Methods}
\begin{table*}[h]
  \caption{Participants Information}
  \label{tab:Participants}
  \begin{tabular}{cl}
    \toprule
    Demographic information &Participant counts\\
    \midrule
    \texttt Age&20-25 years (3), 25-30 years (7), 30-35 years (9), 35-40 years (5), >= 40 years (2)\\
    \texttt Gender&Male(24), Female(2) \\
    \texttt Education level&Middle school(2), High school(12), Junior college(7), Bachelor's (5)\\
    \texttt Years of being safety drivers&<1 year (4), 1-3 years (12), 3-5 years (8), >5 years (2)\\
    \texttt AV technologies worked with&L3 (5), L4 (12), L3 and L4 (9)\\
    \texttt Number of AV companies worked in& 1 company (15), 2 companies (9), 3 companies (1), >3 companies (1)\\
    \texttt Employment Status&Employed (17), Dimission (9)\\
    \bottomrule
  \end{tabular}
\end{table*}

In order to gain a deep understanding of safety drivers’ work practices and experiences, we conducted semi-structured interviews with 26 safety drivers in China from March to July 2022, utilizing further in-depth probing and detailed inquiry~\cite{rubin2011qualitative}. All the interviews were conducted remotely due to the COVID-19 pandemic. This study was approved by the author's organization's ethics committee.

\textbf{\textit{Participants Recruitment}}. 
We recruited participants through professional communities, social platforms, and personal contacts, using snowball and purposeful sampling~\cite{palinkas2015purposeful}. The sampling process was iterative until saturation was reached. 
To gain a more comprehensive picture of the experiences of safety drivers, we intentionally oversampled safety drivers from different companies and female safety drivers that we would not have otherwise. 
Each participant was given a compensation of 30 USD as a token of appreciation.
We recruited 26 participants. Their ages ranged from 25 to 44 years old. 24 were male, and 2 were female. Their years as safety drivers ranged from 0.5 years to 5.5 years. They came from 8 AV companies in China, and 11 of them had work experience at multiple companies.

All participants had experience with highly autonomous vehicles of L3 or L4. According to the interviews, both L3 and L4 AV systems were able to perform most driving tasks on urban roads and operate independently.
When encountering a situation challenging to handle, the L3 AV system would actively hand over control to the safety driver, whereas the L4 AV system would enter a safe state, as defined by the system (e.g., pull over), and would not transfer control to the human actively instead of waiting for the driver to take over passively. 
Since neither L3 nor L4 autonomous vehicles were able to perfectly handle all situations on public roads, it was the primary responsibility of the safety driver to assess the risk level of the current driving scenario in real-time, determine whether the autonomous vehicle could handle it alone, and actively take over control of the vehicle before the risk occurred.

Table~\ref{tab:Participants} presents detailed demographic information for each interviewee.
To preserve an additional layer of anonymity and prevent participants from being identified within their workplaces, we omitted information about their companies and blurred their exact age and years of working experience.

\textbf{\textit{Procedures.}}
Before the interview, two authors conducted a field visit and observed safety drivers' workflow. 
This served the purpose of familiarizing the authors with the safety drivers' work practices, collecting background information, and guiding the outline of the semi-structured interviews.
The key sections of the interview included: (1) participant backgrounds; (2) understanding their work practices; (3) understanding their perceptions of AV; (4) understanding their partnerships with AV; and (5) understanding their working experiences and well-being.
The average duration of each interview was approximately one hour. Each participant was interviewed individually by two researchers using online voice communication software, and all interviews were audio-recorded.

\textbf{\textit{Data Analysis}}.
Our data consisted of 29 hours of audio recordings . Firstly, three authors transcribed the interviews verbatim and examined the transcripts. We then adopted a thematic analysis approach\cite{maguire2017doing} to analyze the 26 transcripts. Each participant's data was qualitatively coded by two or three researchers\cite{mcdonald2019reliability}. The authors collaboratively analyzed the codes and grouped them into themes, and refined the relevant themes in relation to the research questions. At the end, we categorized our resultant codes and themes into three main sections to reflect our findings: (1) work practices of safety drivers; (2) empirical information about the perceptions and partnerships of safe drivers with AV; and (3) safety drivers' work experiences and well-being in working for AV.

\textbf{\textit{Research ethics}}. 
Before commencing the work, this study was approved by the ethics committee of the authors' organization. 
We obtained informed consent from each participant, and participants had the option to decline to answer any questions and terminate the interview at any time. To preserve participant confidentiality, all personally identifiable information was removed from research files, and any identifying details were omitted when quoting participants, due to the sensitive nature of the research.

\section{Results}
This section presents the results of our interviews with safety drivers. Section 5.1 describes the current work practices of safety drivers. Section 5.2 focuses on the human-vehicle partnerships and presents how safety drivers perceive, understand, and work with autonomous vehicles. Section 5.3 describes safety drivers' experiences, well-being, and challenges in working for autonomous vehicle industries.

\subsection{Being a safety-driver}

\textbf{\textit{Recruitments}}.
Self-driving companies generally hire experienced drivers to ensure that autonomous vehicles can complete driving tasks safely and are corrected in time to avoid risks and accidents.
According to our qualitative data, all participants' companies valued extensive driving experience, excellent responsiveness, and good driving habits, which were considered the prerequisites for becoming a safety driver and the basic recruitment requirements.
Most of the safety drivers had prior experience in one or more driving-related occupations, such as taxi drivers (N=8), driving instructors (N=2), chauffeurs (N=2), truck drivers (N=3), and freelance drivers on gig platforms (N=16).
We also performed statistics on the total number of years of the professional driving experience (all professional experience related to driving, including working as full-time drivers, freelance drivers, and safety drivers) that each participant had up to the time of the interview, and the 26 participants had an average of 6.8 years of professional driving experience. 
Additionally, the majority of the safety drivers interviewed (N=24) were under the age of 40.
Our participants responded that AV companies tended to hire younger drivers than older drivers if both meet the recruitment criteria. Although, we did not find solid facts and statistics to prove that younger safety drivers are better equipped to supervise autonomous vehicles compared to older drivers in this study, the bias and stereotypes surrounding older workers in the labor market may result in AV companies preferring to hire younger safety drivers.

\begin{quote}
\textit{"Safety drivers need to understand some basic AV algorithmic knowledge and technical principles, but older people are often seen as being less open to new things and slower to learn. I guess that's why our company would rather recruit and train younger drivers than older drivers."} (P4)
\end{quote}

\begin{quote}
\textit{"You must have very sharp reflexes to monitor self-driving cars. Older safety drivers may not react as quickly as younger ones."} (P17)
\end{quote}

\begin{quote}
\textit{"Working as a safety driver is also physically demanding. You need to be stuck in the car all day and maintain a high level of concentration for a long time. That takes a lot of energy and physical strength, and the younger may be more preferable to the older."} (P2)
\end{quote}

\textbf{\textit{Motivations}}. 
Our research showed that safety drivers are usually low-income workers who depend on their driving skills but have little knowledge about autonomous vehicles.
Similar to other blue-collar groups in developing countries, they often have limited access to employment and education~\cite{mcguinness2008characteristics,strobl2002large}.
Only seven of our participants had a bachelor's degree.
They relied on their driving skills and had few other options for making a living.
Many participants expressed that being a safety driver was a worthwhile opportunity, and a sound choice for them, as it could provide a reasonable and stable salary and improve their standard of living to some extent. As P8 said:

\begin{quote}
\textit{"I'm not sure what else I can do besides driving. Although working as a safety driver doesn't pay very well, it's enough to cover my needs. The company offers me comprehensive insurance and a housing fund, which is much better for me than driving on DIDI (a gig driver platform in China)."} 
\end{quote}

Aside from the financial benefits, one of the main reasons they wanted to be safety drivers was their fascinations with and enthusiasms for AV.
According to our interviews, most of the participants (N=24) did not have technological backgrounds.
They had little AV knowledge but a lot of enthusiasms for AV industries.
Consistent with studies of low-skilled workers in AI industries, they were more likely to be drawn to these industries by the promising future of emerging technologies~\cite{wang2022whose,turner2021human}. 

\begin{quote}
\textit{"I love trying new things, so the chance to work with autonomous driving really gets me pumped!"} (P2)
\end{quote}

\begin{quote}
\textit{"During the interview, the recruiter introduced the development mileage of autonomous driving to me. I thought it was amazing and wanted to get involved to witness its transformation. It would be a great opportunity to improve myself and learn some new skills."} (P12)
\end{quote}

\textbf{\textit{Training and assessment}}. 
Meeting the recruitment criteria did not guarantee these workers would become safety drivers eventually.
Before their employment confirmation get approved, they must undergo rigorous training and pass several rounds of assessments.
According to our participants, the training mainly focused on how to operate an autonomous vehicle safely and driving behavior norms, through theoretical and practical training and lasted a few weeks to months depending on their companies.
The theoretical training typically covered fundamental AV knowledge, driving behavior codes, AV control methods, AV supervision precautions, accident treatment, etc.
Most participants reported that their companies did not provide them with in-depth information regarding the workings of self-driving technologies.
P15 mentioned: 
\begin{quote}
\textit{"In the company's perspective, it is enough for us to drive the car safely, and we do not need to know the technologies behind it very well. Also, since we aren't particularly skilled at learning technology, the company doesn't feel it's worth the effort to teach us."} 
\end{quote}

For practical training, novice safety drivers typically operated AV under the supervision of coaches or senior safety drivers. The training scenarios progressed from simple proving grounds to complex public urban roads.
Over several weeks, they developed their supervisory skills and became familiar with AV operation. There were multiple rounds of evaluation, ranging from theoretical exams to practical operations, with a focus on responsiveness and driving behavior.
Those who didn't pass the test must be retrained or leave the job. \textit{"We had to pass three exams in total, and these exams were very strict. About half of the people in our group failed,"} P15 said. Workers who successfully pass those assessments can eventually become safety drivers, starting their journeys to working with autonomous vehicles.

\textbf{\textit{Responsibilities}}. 
Safety drivers were responsible for supervising and operating AV to finish road testing tasks and ensuring the vehicles' safety and conformity.
According to our interviews, takeover, which typically includes braking, throttling up, and turning the steering wheel, was the most crucial operation for safety drivers to interfere with AV, and important not only for vehicles' safety but also for the research and development team to analyze system defects according to its records. 
Safety drivers should ensure both the safety of the vehicle and the reliability of the data produced by takeover. 
Hence, their takeover decision-making needs to strike a balance between safety and data quality. To achieve this, they need to form precise mental models to predict the behavior of autonomous vehicles and acquire accurate situational awareness of the driving environment.
There were two types of road testing forms: "1 safety driver + 1 AV" and "1 safety driver + 1 AV + engineer(s)."
Additionally, based on the different company's organizational structure, division of responsibilities, and tasks assigned, safety drivers might be required to perform other duties such as data recording, debugging assistance, and hardware maintenance.

\textbf{\textit{Performance Appraisal}}. 
The performance appraisal systems and responsibility allocation regulations for safety drivers differed depending on the company's philosophy and policies. Common criteria for evaluating their performance included mileage driven, working hours, and accident rate. The accident rate was the most important criterion and was taken very seriously by AV companies. 8 participants reported that they would face punishments or even be fired by their companies for accidents.

\begin{figure*}
  \centering
  \includegraphics[width=\textwidth]{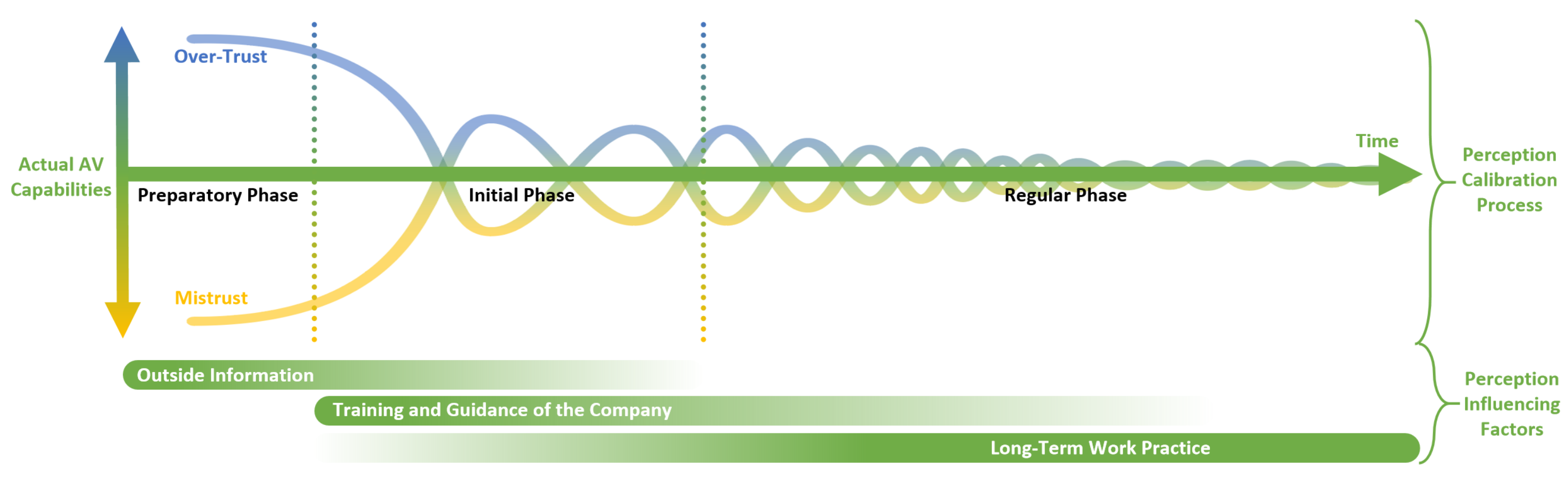}
  \caption{The calibration process of safety drivers' perceptions of autonomous vehicles' capabilities.}
  \Description{This figure presents the calibration process of safety drivers' perception to autonomous vehicle system. The calibration process is divided into three phases: preparatory phase, initial phase, and regular phase. There are 3 perception influence factors: outside information, training and guidance of company and long-term work practice.The preparatory phase is influenced by outside information and this factor decreases over time. The initial initial phase and regular phase are influenced by training and guidance of company and long-term work practice. The factor of training and guidance of company decreases over time. The factor of work practice increases over time.}
\end{figure*}

\subsection{Working with AI: How safe drivers perceive, understand, and partner with AV}

\subsubsection{Forming and calibrating perceptions of AV}

Based on the qualitative data, we identified factors that influence driver perceptions of AV capabilities and classified the calibration process into three stages: preparatory, initial, and regular, as shown in Figure 1.
We found that during the preparatory phase, due to a lack of knowledge about AV, their perception was formed based on outside sources such as news media, journalists, social media, etc., which often contain incorrect information.
And there was a higher likelihood of technophobia~\cite{brosnan2002technophobia} and technopraise~\cite{huesemann2011techno} among this group, which may result in over-trust or mistrust about AV capabilities. In the initial phase, as they gain access to AV and receive training and guidance from their companies, their perceptions will be quickly calibrated.

\begin{quote}
\textit{"At first, I was skeptical. I couldn't wrap my head around how a few tons of metal could drive on its own. I was even scared to get into the car at the start of my work. But after a few days, I started to feel more comfortable, and it exceeded my expectations."} (P10)
\end{quote}

\begin{quote}
\textit{"I used to think self-driving cars were all they were cracked up to be, but after driving it during training, I found that it wasn't like what I had seen on TV. I found that it wasn't like what I had seen on TV. I just couldn't relax and trust it to drive itself."} (P26)
\end{quote}

In the regular phase, safety drivers continuously calibrated their perceptions through their work practices, causing their mental models to converge with the actual capabilities of AV over time. \textit{"It took me about a year to really get to know the car, and I feel like I'm getting better at making predictions now" }, P1 said. 

We found that company-level factors such as training and guidance had a significant impact on safety drivers' perceptions in the initial phase. However, in the regular phase, their AV perceptions became more based on their hands-on practices than company-level factors.
Many participants reported that the theoretical pipe-lined training they received didn't help much with calibrating their mental models, and it was difficult to apply that theoretical knowledge into real-world driving practices, especially in high-risk and emergency situations. Instead, they found that constantly trialing and erroring in their work practices was a better way for them to explore the boundaries of AV capabilities and calibrate their perceptions accurately. Also, they reported that the lessons gained in these processes would be more impressive.

\begin{quote}
\textit{"Although the company has informed us that the radar often misses low objects, if you suddenly encounter obstacles in the road, it's tough to act on that information right then and there. But after being startled, you can remember this lesson very well and handle it better next time."} (P12)
\end{quote}

Additionally, participants reported that after the "novice period," their companies usually paid less attention to the accuracy of their mental models about AV and didn't provide adequate training and assessments in a timely manner. There was also a lack of official methods for continuously calibrating their perceptions while working with AV on a long-term basis.
With each update to the algorithms, system iteration, or hardware change, safety drivers had to form new mental models. However, they didn't know what the "standard answer" was. They had to test and adjust their perceptions in high-risk environments, which resulted in their being potentially exposed to safety risks resulting from misperceptions, and they had to assume responsibilities for those risks.

\begin{quote}
\textit{"The engineers usually only  give us a rough idea of version characteristics when the version is updated, and they don't know when the car might have a malfunction. We must concentrate our attention while driving to try and identify any potential problems with the car."} (P22)
\end{quote}

\subsubsection{Takeover in real world}
\begin{figure*}
  \centering
  \includegraphics[width=\textwidth]{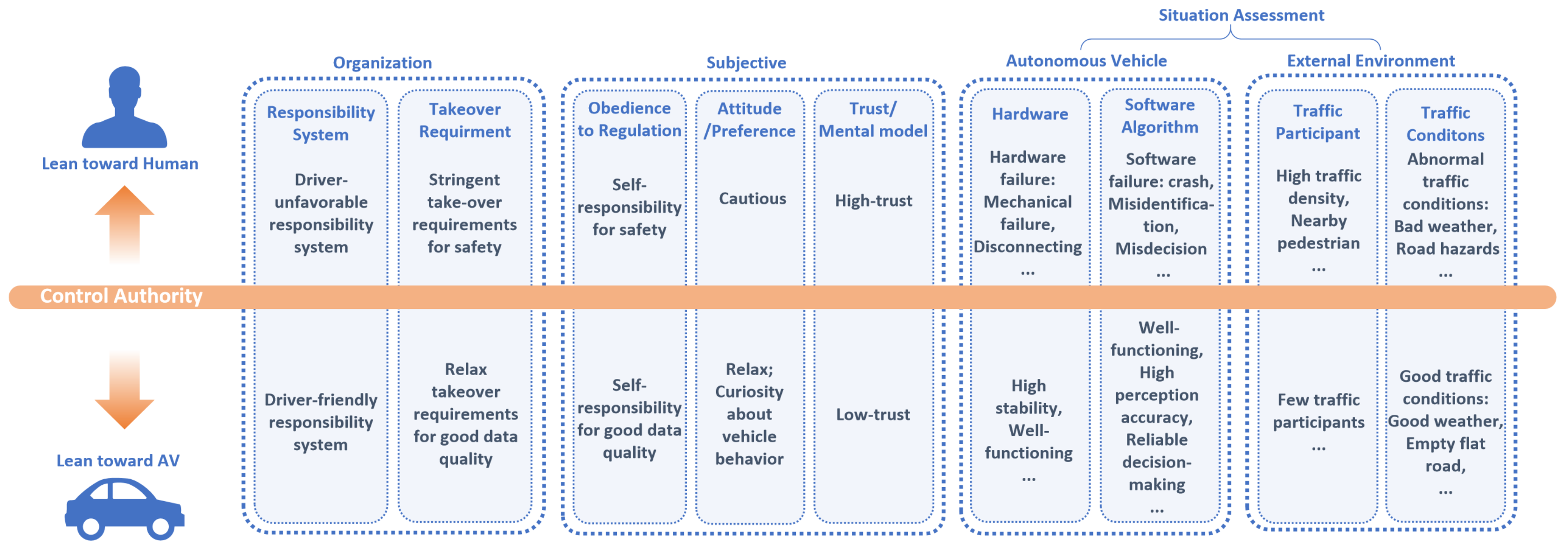}
  \caption{Influencing factors of control authority between the human driver and the autonomous vehicle.}
  \Description{This figure presents the influencing factors of control between safety drivers and autonomous vehicles. There are four sets of factors: organization, subjective, autonomous vehicle and external environment. The following factors will influence the leaning of control authority toward the human driver: driver-unfavorable responsibility system and stringent take-over requirements for safety in organization factor set; self-responsibility of safety, cautious attitude, and high-trust in subjective factor set; hardware failure including mechanical failure and disconnecting, and software failure including crash, misidentification and misdecision in autonomous vehicle factor set; high traffic density, nearby pedestrian, abnormal driving conditions including bad weather and road hazards in external environment factor set. The following factors will influence the leaning of control authority toward the autonomous vehicles: driver-friendly responsibility system and relax take-over requirements for good data quality in organization factor set; self-responsibility of good data quality, relax attitude, curiosity about vehicle behavior, and low-trust in subjective factor set; strong stability, well-functioning, high recognition accuracy, high decision ability Well-functioning in autonomous vehicle factor set; few traffic participants nearby, good driving conditions in external environment factor set.}
\end{figure*}

According to the interviews, takeover was the most important operation for safety drivers to interfere with AV, including braking, throttleing up, and turning the steering wheel, which often took place in risky and unexpected situations and would be recorded by the AV system for problem analysis. 
For the majority of the systems our participants worked with (N=23), once the driver took control of the vehicle, the automation system would shut down and transfer control to the human. 

From the qualitative data, we identified three sets of factors that impacted safety drivers' takeover decision-making and vehicle control authority: organizational factors, personal tendencies, and real-time situations (including the autonomous vehicle and the external environment), as shown in Figure 2.

We found that in low-risk and non-emergency situations, the safety drivers were more likely to be influenced by company factors when making takeover decisions. But in emergency situations, participants expressed a desire to take control of the vehicle and made more subjective and intuitive takeover decisions, which may go against the company's requirements, guidelines, and advocacy.

\textbf{\textit{Organizational tendencies vs. individual tendencies.}} 
Many participants reported that their companies usually guide the driver's takeover criteria based on corporate strategies.
Companies that prioritized efficiency in testing and the quality of takeover data were more likely to ask their safety drivers to relax their takeover criteria.
P23 mentioned,

\begin{quote}
\textit{"My company wants us to be less sensitive to takeovers so we can better know the reaction of AV. If you feel something wrong and take over it immediately, you won't know
what its abnormal reaction exactly is."} 
\end{quote}

While companies that prioritized test safety were more likely to ask their safety drivers to take over in advance. P19 said,

\begin{quote}
\textit{"My leader always emphasizes, 'if you feel something wrong, you must take-over promptly, and you should not tempt the reaction boundaries of the car, otherwise, you are risking the company's property and your personal safety,' so we rarely tempt the limits of the car." } 
\end{quote}

Moreover, safety drivers' takeover attitudes also fluctuated according to their subjective tendencies. P24 said,

\begin{quote}
\textit{"Occasionally, the car will act strangely, and if there are no other cars on the road, I will not take over control purposefully to see how the car will behave. Obviously, I shouldn't do that, and my actions may be recorded by the system. I only do this on a few rare occasions out of curiosity."}
\end{quote}

Sometimes, there were conflicts between the company's tendencies and individual tendencies. In non-emergency situations, these tensions were not obvious. Most safety drivers tended to follow their companies' guidelines to adjust their takeover attitudes and tighten or relax their takeover criteria. As P8 mentioned, \textit{"This is my responsibility to meet my company's requirements. I need to get my job done well." } 

However, in high-risk emergencies, participants reported that they would disregard the company's guidelines and rely on their intuition due to concerns about the consequences of risk, such as their personal safety, liability for accidents, and traffic laws. They explained that they wanted to take control of the vehicle and felt that having more control would give them a greater sense of safety.

\begin{quote}
\textit{"If I feel danger, I will take over control. In that moment, I am unable to give any thought to the company's requirements."} (P12)
\end{quote}

\begin{quote}
\textit{"After taking over and getting the wheel in my hands, I feel relieved and safer."} (P10) 
\end{quote}

Some participants said they had more faith in their own decisions and believed that human decisions were better than those made by machines, especially in emergency situations.
It is worth mentioning that the majority of the participants (N=21) stated that they preferred to take control of the vehicle themselves in risky situations, even if the self-driving decision might be the correct one. As P16 stated,

\begin{quote}
\textit{"Even if it happens many times that the autopilot is right and my judgment is wrong, the next time there is a conflict between our decisions, I would still choose to make it listen to me.
Although its ability is excellent, I don't think I can rest assured to give all of myself to it."}
\end{quote}

\textbf{\textit{For real-time situation assessment.}} 
Most safety drivers (N=24) reported they needed to pay much more attention when supervising the self-driving vehicles compared to when they were driving manually. Extensive lab-experiment-based literature describes that high-level autonomous driving may bring distraction~\cite{cunningham2015autonomous}, loss of attention~\cite{mok2015emergency}, decreased situation awareness~\cite{gibson2016situation}, etc. to drivers. However, few participants mentioned that their level of attention decreased when supervising AV, even for long periods of time. There were even some reported that their level of attention increased with driving time. 

\begin{quote}
\textit{"There are so many unpredictable factors involved. When driving for a long time, I feel that my focus improves."}  (P22)
\end{quote}

\begin{quote}
\textit{"The more time you spend driving, the more tense you become."}   (P1)
\end{quote}

Safety drivers were required to take regular breaks, and their driving time typically did not exceed one hour at a time. This was perhaps one reason why there were no reports of decreased attention caused by long-time automation. Also, participants interviewed in this study were urban road safety drivers, who encountered driving scenarios usually with low monotony and high uncertainty. 
They had to deal with the uncertainty of self-driving hardware, software, and external environments and making quick decisions when something wrong. 
The unpredictability factors in the real world made them continuous high attentions. 
\textit{"I need to pay more attention to the surroundings and detect them more frequently when monitoring AV. I will be very alert and my attention will be more focused,"} P7 said. Meanwhile, long periods of concentration also lead to fatigue in safety drivers' work. P2 said, \textit{"It's very brain-intensive, and I feel very tired at the end of the day." }  

Moreover, participants reported that, due to the fact that their personal safety and work performance were closely tied to the safety of autonomous cars in the real world, they were unable to let their guard down, which was different from the safety-guaranteed laboratory-based studies. Additionally, upon the company's request, the perception and decision-making of safety drivers needed to be paralleled with that of autonomous vehicles. Hence, reducing attention and handing over part of the driving perception tasks to AV would have been negligent for safety drivers. We inferred that one of the possible reasons for increased attention was that our participants were professional safety drivers, had passed related training, and would increase their attention proactively at work. But we did not exclude the possibility that the sensitivity of their work might have deterred them from speaking more critically and admitting their dereliction of duty.

\subsubsection{Cognitive Preferences and Characteristics of AV Technologies}

\textbf{\textit{Know little but want to know more.}} 
The close interaction between safety drivers and AV made exposure to AV technologies in their daily work inevitable, such as learning codes to assist engineers in testing, detecting, and fixing minor AV problems, which reduced their sense of separation and awe about the technology while also stimulating their curiosity. Despite the fact that most our participants had limited knowledge of AV technologies and no technical background, they expressed a desire to learn about how AV works and AV technical principles. P14 said,

\begin{quote}
\textit{“I spend every day with this car, but I don't know what it's thinking. I want to know why it makes certain decisions and behaves the way it does, and I want to know the logic behind it.”} 
\end{quote}

\textbf{\textit{Learning from interacting with engineers.}}
We found that most AV companies believed that it was sufficient for safety drivers just to be drivers for AV and did not invest in technology-related training for them.
Participants reported that most of their understanding of AV technologies gained through their interactions with engineers. P13 mentioned, 

\begin{quote}
\textit{"When testing with the engineer, he is too busy to handle all the tasks himself. He usually teaches me some codes so that I can assist him in tuning the program, and show me how to query the database and fix some bugs. I often ask him why certain things can be tuned in certain ways during the test, and he will explain what the code stands for and what else it can do."} 
\end{quote}

P17 mentioned that he would establish a good relationship with engineers who were willing to share their knowledge, in order to facilitate future consultation and learning.

\begin{quote}
\textit{"If I build a good relationship with the engineer, he may request that I be paired with the captain so that I can spend more time communicating with him and learn something."} 
\end{quote}

Although communication with engineers can help safety drivers understand technology to a certain extent, this knowledge transfer was limited, and the information obtained might be incomplete. 
Engineers were not always concerned with how much safety drivers had mastered or whether their knowledge mastered was accurate. \textit{"The company will not teach us these, and the engineers only speak these to us briefly, so we can only understand the superficial aspects," }  P15 said.

\textbf{\textit{Invisible algorithms and visible hardware.}} 
We found that most safety drivers tend to associate invisible algorithms with visible hardware. When we asked them about their understanding of AV, most answered by talking about their knowledge of AV hardware. P19 said,

\begin{quote}
\textit{"I don't know much about software algorithms, but I do understand some basic hardware concepts like lidar, millimeter-wave radar, cameras, and so on. I understand their perception range, parameters, and simple hardware debugging methods."} 
\end{quote}

We also found they would naturally associate the level of AV capabilities with the vehicle's hardware systems. When there was a problem with the AV, they would first assume it was a hardware issue, and then consider the possibility of an algorithm problem. \textit{"If the car suddenly stalls, I will first check its radar, camera, and other sensors,"} P15 said.
Participants also reported their greater interests in understanding hardware knowledge than software algorithms. They believed hardware knowledge was easier to comprehend and more useful in their daily work. P14 and P16 mentioned that they would like more training on hardware knowledge from the company. P10, P18, and P21 expressed their desire to switch to a career as hardware engineers.

\begin{quote}
\textit{"I prefer purely mechanical things."} (P14)
\end{quote}

\begin{quote}
\textit{"I have difficulty understanding those red and green codes, but I'm more familiar with how the hardware works."} (P16)
\end{quote}

\subsection{Working for AI: Safety Drivers' Experiences and Well-being}

\subsubsection{Work Experiences}

\textbf{\textit{From technology experiences to work experiences.}} 

Safety drivers need to work in close contact with AV in their daily tasks. In such a work context, they were the supervisor and also the experienced user of AV, and their happiness and sadness at workplaces were closely linked to the AV they worked for.

As supervisors, participants reported that they were able to gain a great sense of effectiveness when the issues they reported were resolved, and they could feel firsthand the progress of AV.
\textit{"When we report a problem and it is suddenly resolved after a few days, I feel a sense of satisfaction knowing that I was able to contribute to the improvement of the vehicle,"} P24 said. Some indicated that they feel neglected and unappreciated when they do not receive feedback or the issues they report are not resolved.
\textit{"We've reported this issue several times, but it still happens as soon as the car approaches this intersection. It's been a long time, and the problem has not been resolved. I feel like they just don't want to pay much attention to me,"} P15 said.

As users of AV, participants reported that they have positive experiences when the vehicles behave in accordance with human thoughts in their daily driving.

\begin{quote}
\textit{"I want it to go faster, and it does; I want it to slow down, and it does. It's as if it knows what you're thinking, and it feels like I've become integrated with it."} (P11)
\end{quote}

Some indicated that the inconsistency between the vehicle's behavior and the human's intentions and expectations might trigger their negative experiences.

\begin{quote}
\textit{"It has its own thoughts, and I have mine. Sometimes it doesn't inform me, it just goes right by and it just lets me agitated."} (P21)
\end{quote}

Some expressed disappointment that the limitations of algorithms sometimes override human intentions. \textit{"I like the aggressive driving style, but the car just drives very conservatively,"} P23 said. Participants also mentioned they would prefer to work with the "perfect machine".

\begin{quote}
\textit{"I've worked for two different companies. The technology of the first company was not very developed, as the autonomous vehicle would frequently brake abruptly and cause conflicts with other road participants. When I finish my daily work, I'm often in a bad mood. But now, the technology of my current company is much more developed, and the autonomous vehicle is much smoother and more comfortable to ride. I feel much better now."} (P9)
\end{quote}

\textbf{\textit{Technology mediates social interactions in the workplace.}}
Autonomous driving technology also had an impact on the social interactions among safety drivers at workplaces. We found that AV technologies improved communication and collegiality among them. P9 mentioned that when she first started at the company, she was unfamiliar with everything. AV became an ice-breaker topic between her and her colleagues, hastening their acquaintance.
Safety drivers were willing to discuss their guesses about AV with others and share problem solutions, which gave them a sense of accomplishment and self-efficacy and strengthened their colleague relationships. \textit{"We often discuss common problems together, such as navigation and perception issues, and we are willing to help each other,"} P11 said.

We also found that safety drivers' interactions with engineers significantly enhanced their desire to learn more about AV technologies and their self-efficacy, which in turn impacted their career planning.

\begin{quote}
\textit{"The engineers have taught me a lot, and I feel like I can do the same things they do. I'm thinking about becoming an engineer in the future."} (P21)
\end{quote}

While we observed positive effects of technology on workplace interactions, we also found that technology hindered some safety drivers' self-expression to some extent. 
Some participants mentioned that their lack of technological knowledge made them hesitant to express their opinions. They tended to suppress their expression in certain situations because they feared being denied or ridiculed.

\begin{quote}
\textit{"When I encounter some simple problems that I don't know how to solve, I'm afraid to ask my leader because I don't want them to think I'm incompetent. I often ask my colleagues, but they don't always know the answers to my questions."} (P11)
\end{quote}

\begin{quote}
\textit{"Sometimes I have some guesses about AV problems. I usually wait until I'm absolutely certain before sharing them with others. After all, I'm not a professional engineer, so I'm concerned that my ideas may be incorrect."} (P23)
\end{quote}

\textbf{\textit{Real world driving experiences.}} 
When safety drivers drove the "strange-looking black box" on public roads, they attracted more attention from the outside world and had novel experiences with AV, which gave them the pride of being noticed, unnecessary annoyances, and risks caused by other traffic participants' curiosity and low acceptance of new things.
Some participants expressed a very positive feeling about the curiosity and attention of the outside world caused by autonomous driving and felt proud and happy. 

\begin{quote}
\textit{"When I'm waiting for the traffic light at the intersection, people often take pictures of the autonomous vehicle, which gives me the illusion that they are paying attention to me."} (P2)
\end{quote}

Some also mentioned that the outside attention would cause them unnecessary distress.

\begin{quote}
\textit{"Because this vehicle is too conspicuous, the traffic police sometimes specifically spot-check me."} (P6)
\end{quote}

\begin{quote}
\textit{"When the vehicle breaks down on the road, some passers-by may go near and take a look, and even stop to take a picture, which made me very embarrassed."} (P16)
\end{quote}

Furthermore, participants stated that they had to accept the risk of low acceptance of AV by other road participants, such as malicious behavior caused by curiosity.

\begin{quote}
\textit{"There are many cars on the road that often deliberately come to a halt in front of you because they are curious how the self-driving car will react."} (P20)
\end{quote}

\begin{quote}
\textit{"A few years ago, the news media widely reported that AV would replace human drivers. During that time, I was often bullied by taxis while conducting road tests. But it's much better now, they have become accustomed to our vehicles."} (P1)
\end{quote}

Moreover, due to their limited capabilities, autonomous vehicles may not be able to handle all road test conditions perfectly, which may result in exclusion by other road participants. P23 mentioned that his vehicle sometimes has to bear the consequences of mistakes made by other autonomous vehicles while driving:

\begin{quote}
\textit{"The cars in our team often make mistakes on the road and affect others (road participants). Those drivers probably hold a grudge, and the next time they encounter our team's autonomous vehicles, they will deliberately bully them, even if the car they retaliate against is not the same one that disrupted them last time. I am often wrongly accused."}
\end{quote}

\subsubsection{Taking Risks Accumulated from upstream of AV Industry}

\textbf{\textit{Forced to expose themselves to accumulated risks from AV upstream.}}
During road tests, safety drivers, as the downstream workforce of the AV industry, were forced to face the accumulated risks from multiple upstream links, including algorithm development, hardware manufacturing, assembly, etc.
Unlike traditional testers~\cite{ekwoge2017tester}, they needed to verify and test AV in high-risk real-world environments, where a minor omission from other links could expose them to great risks.
Based on the interviews, developers in other upstream links of AV were not required to participate in road tests like safety drivers. In such a workflow, these stakeholders were unable to predict precisely how the AV they developed would perform in the real world, which might reduce their sense of responsibility and increase the likelihood of negligence.

\begin{quote}
\textit{"I remember a remote debugging session. The engineer told me the car had been fixed and let me test it. But when I tested it, I found that the problem had not been set up right, and I almost collided with an obstacle."} (P7)
\end{quote}

\begin{quote}
\textit{"There was a system version update, but the new version had many strange problems, and the engineers didn't know what was going on. We felt very unsafe while driving and had to pay very close attention."} (P14)
\end{quote}

Although it was the duty of the safety drivers to look for defects and intervene in AV to prevent risks, relying solely on the safety drivers to identify accumulated problems and avoid accumulated risks is difficult and immoral.

P3 said, \textit{"There are some unexpected situations where it's too late for you to takeover, and no one can change the outcome in such a short amount of time."}
P7 expressed his concerns about safety, \textit{"Every morning when I go out for work, I pray to come home safely,"} and thoughts about changing jobs.

\subsubsection{Ambiguous Responsibility Assignments and Moral wrinkles}

\textbf{\textit{Responsibility assignments.}}
In addition to bearing the accumulated risks, safety drivers also faced ambiguous and undue assignments of responsibility. 
China's traffic laws did not consider self-driving cars as responsible entities. When safety drivers conducted road tests with AV, they were held legally responsible. Even though the majority of the time the car was being driven by AI and not the safety driver, the safety driver was still held accountable for the vehicle. For traffic violations, even those caused entirely by the self-driving system, the safety driver was still punished as the primary responsible party and their driver's license points would be discounted. A poor driving record would be recorded in the driver's file and might impact their driving qualifications or limit their opportunities for other driving-related jobs.
P8 said, 

\begin{quote}
\textit{"If we receive a citation for breaking traffic laws because of the self-driving car, the company will compensate us with some money. However, there is no way to regain the license points that have been deducted, and we have to accept the loss."} 
\end{quote}

Participants said they do not have to take responsibility for passive accidents, but if the accident was caused by the fault of the safety driver or the autonomous vehicle, they may face consequences such as a warning or penalty from the company. Two-thirds of the safety drivers reported that the company's responsibility allocation system was not fair to them. 

\begin{quote}
\textit{"The company has a zero-tolerance policy for accidents. Because safety drivers are hired for the purpose of ensuring safety. If an accident occurs, it is considered the responsibility of the safety driver."} (P15)
\end{quote}

Some participants reported that their companies implemented a shared responsibility system as a way to warn safety drivers not to have an accident.

\begin{quote}
\textit{"The training of new safety drivers is conducted by old safety drivers, and if the new safety driver causes an accident, the old safety driver will take responsibility for the new safety driver."} (P21)
\end{quote}

It is undeniable that this stringent responsibility system can improve the safety drivers' sense of responsibility at work, but it also infringes on their rights and leads to a negative work experience to some extent. There were also some companies that differentiated the responsibilities between human and autonomous vehicles.

\begin{quote}
\textit{"If accidents happen while on autopilot mode, your responsibility may be reduced, but if you were distracted, fell asleep, drank water, used your phone, etc., that would be your responsibility."} (P25)
\end{quote}
 
However, due to the lack of clarity on the boundaries of responsibility, safety drivers may take on more responsibility than they are supposed to.

\begin{quote}
\textit{"In the moment before an accident, you must have taken control instinctively. It can be difficult to determine whether you were involved in causing the accident or not."} (P17)
\end{quote}

\begin{quote}
\textit{"Although it is said that the company will analyze whether the accident was caused by AV or the safety driver through monitoring data and system records, but then again, if there is an accident and you take over, then you may be responsible for involving the accident. If you don't take over, that means you have not performed the duties of the safety driver, and then you are still responsible."} (P13)
\end{quote}

\textbf{\textit{Operator or passive observer.}}
As a new type of occupation, the current corporate systems, laws, and regulations are not fully equipped to guide safety drivers in all situations, leading to ambiguities in road testing.
As described in the previous section, some companies adopt a safety-driver-friendly allocation of collision responsibility. If a collision occurs while the autonomous system is running without any human interference, the company will assume full responsibility for the autonomous system they developed. However, this regulation raises another question of whether the safety driver should intervene in the AV system at the moment of an upcoming accident. Participants reported that they would not stand by and watch an accident happen to reduce their responsibility. They would instead instinctively take over the autonomous vehicle, even if their actions would not save anything.

\begin{quote}
\textit{"Takeover better than an accident."} (P19)
\end{quote}

\begin{quote}
\textit{"I'll take over even if I can't save it. I can't watch the car go into an accident."} (P11)
\end{quote}

\begin{quote}
\textit{"I wouldn't allow an accident to happen for fear of liability, but I don't know what the rest of my colleagues think."} (P8)
\end{quote}

\begin{quote}
\textit{"Even though not intervening could absolve me of responsibility, there's no guarantee that the company wouldn't hold me accountable if an accident occurs. So I believe it's more important to prevent an accident from happening."} (P16)
\end{quote}

\textbf{\textit{Collision leaded by avoid-collision.}}
According to our interviews, AV companies usually adopt very cautious algorithmic strategies to avoid active collisions. However, these overly conservative driving strategies made self-driving cars slow and easy to stop, increasing the chances of being rear-ended by other vehicles. Participants reported that they have to put in extra effort and take on additional risks because of the shortcomings of AV strategies.

\begin{quote}
\textit{"When the self-driving car falls short, it's up to us safety drivers to fill in the gaps. But sometimes the car's capabilities are just too limited and we are also very helpless. But after all, this is our job."} (P22)
\end{quote}

\begin{quote}
\textit{"The safety driver must pay more attention to the rear of the vehicle to avoid being rear-ended, which requires a high level of attentiveness from the safety driver. Sometimes some novice safety drivers are not able to effectively consider the surroundings, so it is quite prone to accidents."} (P25)
\end{quote}

\subsubsection{"I can see the future of AV, but not mine"}

As described in Section 5.1, safety drivers typically relied on their driving skills for their livelihood and had limited career choices. Although they had the opportunity to be exposed to emerging technologies and witness changes in the autonomous driving industry as safety drivers, it was challenging for them to achieve career growth through this job.

\begin{quote}
\textit{"You can't learn anything just by staring at this car every day. I feel as if I've been extinguished after working for a long period of time as a safety driver."} (P9)
\end{quote}

Participants expressed their concerns about their future developments.

\begin{quote}
\textit{"As long as you can drive, you can be a safety driver. There is nothing irreplaceable. Who knows, I might be laid off one day."} (P5)
\end{quote}

They also reported an age crisis among safety drivers. Some companies set age requirements due to the high levels of endurance, sensitivity and responsiveness required for the job.
Despite many studies~\cite{deo2019looking,huang2020age,kim2017takeover} showed that there was no significant difference in takeover response ability between young and middle-aged people, the inherent impression can also expose this group to age discrimination. 
\textit{"Younger drivers may have better response abilities, and the older drivers may be gradually eliminated from the company,"} P10 said.

Moreover, participants also recognized that AV technology is moving towards unmanned operation, and they feared that the role of safety driver may become an overstaffed position that is vulnerable to being eliminated. P18 said, 

\begin{quote}
\textit{"Safety driver is only a transitional position in AV development. With the progress of AV technology, safety drivers may become obsolete. When that happens, I don't know what the company's plan is, and I am not sure about my future."} 
\end{quote}

P9 expressed his ambivalence between his expectations for the development of AV and his future development as a safety driver: 

\begin{quote}
\textit{"I wish for the technology to progress faster to make my job easier, but I don't want it to develop very well because it may jeopardize my employment."} 
\end{quote}

\section{Discussion}

\subsection{From Human-Vehicle Partnerships to broader Human-AI Partnerships}

\textbf{\textit{More effective training and experience transfer for non-technical lay users.}}
Designing effective training programs and experience transfer strategies can help ensure that the technology is used in a responsible and ethical manner. Proper training of the driver in taking over the vehicle and prompt decision making is imperative for ensuring safety.
According to our interviews, although the process of understanding and learning AV technologies may vary for safety drivers based on their individual preferences, experiences, and circumstances, some common characteristics of experience acquisition have been observed in this study. These include: \textit{practices over theories, try and fail over smooth processes, specific and visible over abstract and invisible, and interactions with colleagues over taught lessons.} 
Safety drivers' traits of learning difficulties with AI technologies and their lack of AI knowledge resemble the majority of end-users of AI systems~\cite{ribera2019can}. 
Their cognitive preferences and learning characteristics during the training process can be extended to a broader range of lay users, informing the design of training strategies that are more aligned with user learning preferences to enhance users' understanding of AI technologies.
Hence, we consider that when introducing AI systems to new users, the training programs are supposed to allow for active involvement, visible feedback, and quick response to ensure such systems can be fully understood and analyzed by the lay user. Some HCI researchers have explored novel interaction strategies to train users and improve their understanding of AI systems, such as combining theory with practice through immersive technologies~\cite{makris2016augmented}, collaborative learning assistant agents~\cite{magnisalis2011adaptive}, interactive training games~\cite{ebnali2019user}, etc. We hope the insights gained from this study regarding the technology learning and skill transfer characteristics of safety drivers can inform the design of AI system training strategies for a wider range of AI users.

\textbf{\textit{Mental model calibration and bidirectional communication.}}
In our study, safety drivers constantly updated their mental models during their working practices. However, these changing mental models were not assessed and calibrated in a timely and effective manner. 
Our participants reported usually having misunderstandings about AV systems, as they lacked a clear understanding of the mechanisms of AV systems. They could only verify their mental models by comparing the results produced by AV systems.
In addition, due to the lack of reliable calibration processes, they have to wait until the consequences aroused by their wrong mental model occur before they can calibrate their wrong mental model. 
Therefore, it is crucial to implement effective strategies that support users in calibrating their mental models timely through their day-to-day interactions with AI systems, for example providing real-time or regular feedback and clear and concise explanations about the mechanisms of the AI systems~\cite{mueller2019explanation}. 
Our study also found that safety drivers struggled with communicating their understanding to the AV system, and the AV system was also unable to comprehend the safety drivers' mental models. 
This limited one-way communication may pose challenges to the partnerships between safety drivers and autonomous vehicles and can increase safety risks.
Recent studies also showed that the ability to facilitate effective communication during interactions is more effective in promoting teamwork than technical capabilities~\cite{duhigg2016google,liang2019implicit}. 
Aside from improving the interpretability and transparency of AV to users~\cite{Lyons2013BeingTA,Gyevnar2022AHM,Shen2020ToEO}, AI systems should be able to evaluate the changes in users' mental models over time through building shared mental models and bidirectional communication between humans and machines~\cite{ososky2012importance,andrews2022role}.

\textbf{\textit{Decision-making in high-risk situations.}}
Trende's~\cite{Trende2019AnII} and Adam's~\cite{MillardBall2016PedestriansAV} studies investigated human decision-making in highly autonomous vehicles in time-critical situations, both of their results showed that subjects tended to accept the automated system's suggestions rather than decisions made by their own. Contrary to their studies, our study shows that in emergency and high-risk situations, 
safety drivers take over the AV more by instinct and intuition than by rational decision making. Most safety drivers report that they could not accurately and quickly obtain the real-time decision logic and results of AV, and in this case, they were more inclined to believe their own decisions. 
Safety drivers tend to rely more on instinct and intuition when taking over autonomous vehicles, as they are unable to accurately and quickly understand the rationale behind AV decision-making and make real-time evaluations. As a result, they prefer to trust their own decisions more.
However, sometimes intrusive decision-making may bring risks. As Villemeur~\cite{villemeur1992assessment} mentioned, the likelihood of a human making an incorrect supervisory decision during a short time frame for unexpected, high-consequence scenarios is close to one hundred percent.
We have analyzed the following questions to be solved:

\textit{How to allow users to accurately assess the accuracy of AI decisions.} 
In this study, humans have the absolute power to interfere in AI decisions, while wrong interventions may lead to serious consequences. Therefore, it is crucial to enable safety drivers to accurately assess the AI decisions and make a more informed and responsible control transfer.

\textit{How to evaluate and predict which decision-making, either by humans or AI, is better for an event that did not occur.} Balance decision-making between AI systems and humans has remained a challenge in human-AI collaboration research. Researchers are actively exploring potential solutions, such as establish a third-party decision-making mechanism or evaluation mechanism between these two decision makers, which may be helpful to provide humans with evaluation information about both decision makers~\cite{crompton2021decision,bader2019algorithmic}.

\textit{How to get humans to overcome their limitations (e.g., uncontrollable instinctive reactions, overconfidence) and willingly hand over control to AI, once the AI system is accurate in its decisions and humans have realized that AI is accurate.} 
Collaborative decision-making between humans and AI systems is becoming more widespread. Further exploration is needed to enable users to accurately understand the AI system and overcome any interaction barriers that may arise due to human limitations.

\subsection{Towards Responsible AI: From Safety Drivers' Perspectives}

\textbf{\textit{Support the upstream AV industry's "consequence awareness".}}
The creators of the upper echelons of technology often lack awareness of what the potential consequences of the technology they create may have~\cite{coeckelbergh2020artificial}. 
This can lead to a reduction in the level of caution and responsibility among various stakeholders, which increases the likelihood of creating a "crazy machine"~\cite{Sunstein2018AlgorithmsCB}.
In this study, safety drivers, seen as the testers in the last piece, are required to drive AV on real roads to find defects. They have a better understanding of how the technology performs in the real world than upstream technology developers and are expected to identify accumulated defects and take responsibility for accumulated risks. It is crucial for stakeholders in the upstream autonomous vehicle industry to be aware of the potential consequences of the technology they are involved in. Organizations should also be responsible for embedding responsibility in all stages of technology production, especially those that are often neglected.

\textbf{\textit{Challenges in work practices.}}
Every aspect of the AI industry is closely interconnected, from development, testing, to implementation. 
The bottom-tier AI workers still play a critical role in the AI industry, and many studies have highlighted the significant issues that arise from neglecting those workers~\cite{sambasivan2021everyone,muller2021designing}. We found similar challenges in safety drivers' work practices. 

It is paradoxical and worrying that safety drivers generally have a low knowledge of AV, but their job requires them to be able to predict AV accurately. According to our interviews, companies were more concerned with having safety drivers learn how to operate the machines and less concerned with their understanding of technological mechanisms. Often, safety drivers were only able to acquire AV knowledge through limited learning approaches, such as social media or informal communication with colleagues, and the knowledge acquired through these tracks is often incomplete and inaccurate.
Lack of knowledge of technology hinders their work practices and can lead to misprediction~\cite{mirnig2016framework}, misattribution~\cite{woods1994behind}, etc., which may lead to risks. We suggest that companies should value safety drivers' knowledge of AV and provide them with accessible learning sources. 

Our findings also show that it can be challenging for safety drivers to transfer their training experience with autonomous vehicles to real-world road tests. The calibration of their knowledge and mental model of AV systems relies more on trial and error through hands-on practices than on training. However, this means they must bear real safety and violation risks in order to gain experience, which is a costly way to learn.

Although previous studies~\cite{Walch2016TowardsCD,Dixit2016AutonomousVD} have explored how to help users gain AV operating experience, most of them are based on laboratory and simulated environments. There is limited research on how to effectively transfer this experience to the real world. Therefore, finding more effective training methods to reduce the cost of transferring skills for safety drivers and enable them to form accurate mental models without taking risks is a crucial challenge that needs to be addressed.

Additionally, it was reported that most AV companies focused more on safety drivers' initial training and neglected to provide adequate ongoing training and assessment during safety drivers' extended periods of work. It is recommended that organizations provide long-term mentorship for autonomous vehicle operators, so that they can effectively evaluate the performance and weaknesses of the system, especially as technology evolves. 

\textbf{\textit{Involving front-line AI workers into human-centered AI research.}}
Despite being the lowest-level testers in the AV industry and not directly involved in the optimization of AV, safety drivers have unique, first-hand experiences with autonomous vehicles as front-line workers who have been working closely with them for an extended period of time. 
They often develop their own understanding of AV flaws that may go unnoticed by the research and development team.
Their opinions may assist the research and development team in identifying and correcting defects that have not yet triggered safety risks, and addressing them prior to the occurrence of risks to improve the safety of AV testing.
However, according to our research, safety drivers, as marginal workers, do not have a strong voice and are frequently overlooked for the concerns they report. 
We recommend that the information feedback mechanism be improved and that the opinions and input of front-line automation workers be valued and taken into consideration.
Moreover, since front-line workers like safety drivers have hands-on experience with AI, involving these workers in the technology development process would contribute to the advancement of the technology and lead to a more human-centered approach in AI research. Their perspectives and experiences could also be involved in the development, data collection, and analysis phases to help identify any potential issues or concerns with the AI system and ensure its usability and reliability.

\subsection{Experience Migration: Spy on the Future User Experience of Automation Systems}
As automation advancements progress, the integration of technology systems with human capabilities is creating novel driving experiences~\cite{wiegand2020joy}. 
As the "first passenger of AV," safety drivers' experience of high-level autonomous driving can help us understand the automation experience of a wider group. Through interviews, we learned about the positive and negative aspects of their experience, which were closely related to AV technologies. Additionally, we found that their social interactions with people in workplaces were facilitated by technologies, and exposure to emerging technologies increased their self-efficacy and desire for expression.
We also observe a dynamic and complex change in the human-machine relationship, resulting from complex environments, different AI system qualities, and various personal preferences and personality traits. For example, some drivers felt a strong connection and externalized themselves to AV, whereas some felt weak ties to it. When facing external accusations caused by the AV, some empathized with the car, while some felt embarrassed and tried to cover up its deficiencies by takeover. Some even crossed the line by taking over the car without complying with company rules due to personal preferences or social pressure, and some chose to disassociate themselves with AV by gesturing or mouthing to other road participants when AV made mistakes. 
We learned their real-road driving experiences, which made them experience the pride of being noticed, unnecessary annoyances, and risks caused by outside curiosity and low acceptance.
These safety drivers' experience can help us gain insight into the future user experience of broader AI technologies and automation systems.

\section{Limitations}
We acknowledge that our study is an initial investigation. We only interviewed a small sample size of safety drivers, and the sensitivity of their work may have deterred them from speaking more critically about their experiences, which may lead to subjective bias and a lack of generalizability for our results. To further work on this study, more information would be valuable and could be gathered by conducting more interviews with safety drivers and other stakeholders in the AV industry, for a more complete view. On that basis, more quantitative and confirmatory research could be conducted to validate the findings.

\section{Conclusion}
In this paper, we present how safety drivers perceive, understand, and partner with AV in the real world. We also found that, as front-line workers, safety drivers are forced to take risks accumulated from the AV industry upstream and are also confronting restricted self-development in working for AV development. We discussed the opportunities for human-vehicle partnerships and the improvement of workers' experiences. 
We compared our findings with previous literature and find the gaps between human-AV interactions in  controlled experiment environment and real world long term practices. 
We contribute the first empirical evidence of the lived experience of safety drivers—the first passengers in the development of AV, as well as the grassroots workers for AV. We hope that this paper will provide valuable insights for more concerted and confirmatory research efforts in the area and consequently contribute toward the implementation of automated driving. 

\begin{acks}
This project is supported by the National Natural Science Foundation Youth Fund 62202267.
We thank the participants who shared their lived experiences with us, as this study would not have been possible without them.
\end{acks}

\bibliographystyle{ACM-Reference-Format}
\bibliography{sample-base}


\begin{thebibliography}{96}


\ifx \showCODEN    \undefined \def \showCODEN     #1{\unskip}     \fi
\ifx \showDOI      \undefined \def \showDOI       #1{#1}\fi
\ifx \showISBNx    \undefined \def \showISBNx     #1{\unskip}     \fi
\ifx \showISBNxiii \undefined \def \showISBNxiii  #1{\unskip}     \fi
\ifx \showISSN     \undefined \def \showISSN      #1{\unskip}     \fi
\ifx \showLCCN     \undefined \def \showLCCN      #1{\unskip}     \fi
\ifx \shownote     \undefined \def \shownote      #1{#1}          \fi
\ifx \showarticletitle \undefined \def \showarticletitle #1{#1}   \fi
\ifx \showURL      \undefined \def \showURL       {\relax}        \fi
\providecommand\bibfield[2]{#2}
\providecommand\bibinfo[2]{#2}
\providecommand\natexlab[1]{#1}
\providecommand\showeprint[2][]{arXiv:#2}

\bibitem[Abu~Bakar et~al\mbox{.}(2022)]%
        {abu2022synthesis}
\bibfield{author}{\bibinfo{person}{Amirul~Ibrahim Abu~Bakar},
  \bibinfo{person}{Mohd~Azman Abas}, \bibinfo{person}{Mohd~Farid Muhamad~Said},
  {and} \bibinfo{person}{Tengku~Azrul Tengku~Azhar}.}
  \bibinfo{year}{2022}\natexlab{}.
\newblock \showarticletitle{Synthesis of Autonomous Vehicle Guideline for
  Public Road-Testing Sustainability}.
\newblock \bibinfo{journal}{\emph{Sustainability}} \bibinfo{volume}{14},
  \bibinfo{number}{3} (\bibinfo{year}{2022}), \bibinfo{pages}{1456}.
\newblock


\bibitem[Altendorf et~al\mbox{.}(2019)]%
        {altendorf2019utility}
\bibfield{author}{\bibinfo{person}{Eugen Altendorf}, \bibinfo{person}{Constanze
  Schreck}, \bibinfo{person}{Gina We{\ss}el}, \bibinfo{person}{Yigiterkut
  Canpolat}, {and} \bibinfo{person}{Frank Flemisch}.}
  \bibinfo{year}{2019}\natexlab{}.
\newblock \showarticletitle{Utility assessment in automated driving for
  cooperative human--machine systems}.
\newblock \bibinfo{journal}{\emph{Cognition, Technology \& Work}}
  \bibinfo{volume}{21}, \bibinfo{number}{4} (\bibinfo{year}{2019}),
  \bibinfo{pages}{607--619}.
\newblock


\bibitem[Andrews et~al\mbox{.}(2022)]%
        {andrews2022role}
\bibfield{author}{\bibinfo{person}{Robert~W Andrews}, \bibinfo{person}{J~Mason
  Lilly}, \bibinfo{person}{Divya Srivastava}, {and} \bibinfo{person}{Karen~M
  Feigh}.} \bibinfo{year}{2022}\natexlab{}.
\newblock \showarticletitle{The role of shared mental models in human-AI teams:
  a theoretical review}.
\newblock \bibinfo{journal}{\emph{Theoretical Issues in Ergonomics Science}}
  (\bibinfo{year}{2022}), \bibinfo{pages}{1--47}.
\newblock


\bibitem[Bader and Kaiser(2019)]%
        {bader2019algorithmic}
\bibfield{author}{\bibinfo{person}{Verena Bader} {and} \bibinfo{person}{Stephan
  Kaiser}.} \bibinfo{year}{2019}\natexlab{}.
\newblock \showarticletitle{Algorithmic decision-making? The user interface and
  its role for human involvement in decisions supported by artificial
  intelligence}.
\newblock \bibinfo{journal}{\emph{Organization}} \bibinfo{volume}{26},
  \bibinfo{number}{5} (\bibinfo{year}{2019}), \bibinfo{pages}{655--672}.
\newblock


\bibitem[Baecker(2019)]%
        {Baecker2019AutomationWA}
\bibfield{author}{\bibinfo{person}{Ronald~M. Baecker}.}
  \bibinfo{year}{2019}\natexlab{}.
\newblock \showarticletitle{Automation, work, and jobs}.
\newblock \bibinfo{journal}{\emph{Computers and Society}}
  (\bibinfo{year}{2019}).
\newblock


\bibitem[Baltrusch et~al\mbox{.}(2022)]%
        {baltrusch2022human}
\bibfield{author}{\bibinfo{person}{SJ Baltrusch}, \bibinfo{person}{F Krause},
  \bibinfo{person}{AW de Vries}, \bibinfo{person}{W van Dijk}, {and}
  \bibinfo{person}{MP de Looze}.} \bibinfo{year}{2022}\natexlab{}.
\newblock \showarticletitle{What about the Human in Human Robot Collaboration?
  A literature review on HRC’s effects on aspects of job quality}.
\newblock \bibinfo{journal}{\emph{Ergonomics}} \bibinfo{volume}{65},
  \bibinfo{number}{5} (\bibinfo{year}{2022}), \bibinfo{pages}{719--740}.
\newblock


\bibitem[Baltzer et~al\mbox{.}(2014)]%
        {baltzer2014mediating}
\bibfield{author}{\bibinfo{person}{Marcel Baltzer}, \bibinfo{person}{Eugen
  Altendorf}, \bibinfo{person}{Sonja Meier}, \bibinfo{person}{Frank Flemisch},
  \bibinfo{person}{N Stanton}, \bibinfo{person}{S Landry}, \bibinfo{person}{GD
  Bucchianico}, {and} \bibinfo{person}{A Vallicelli}.}
  \bibinfo{year}{2014}\natexlab{}.
\newblock \showarticletitle{Mediating the interaction between human and
  automation during the arbitration processes in cooperative guidance and
  control of highly automated vehicles: basic concept and first study}.
\newblock \bibinfo{journal}{\emph{Advances in Human Aspects of Transportation
  Part I}} (\bibinfo{year}{2014}), \bibinfo{pages}{439--450}.
\newblock


\bibitem[Banks et~al\mbox{.}(2018)]%
        {banks2018keeping}
\bibfield{author}{\bibinfo{person}{Victoria Banks}, \bibinfo{person}{Emily
  Shaw}, {and} \bibinfo{person}{David~R Large}.}
  \bibinfo{year}{2018}\natexlab{}.
\newblock \showarticletitle{Keeping the driver in the loop: The
  ‘Other’ethics of automation}. In \bibinfo{booktitle}{\emph{Congress of
  the International Ergonomics Association}}. Springer,
  \bibinfo{pages}{70--79}.
\newblock


\bibitem[Brosnan(2002)]%
        {brosnan2002technophobia}
\bibfield{author}{\bibinfo{person}{Mark~J Brosnan}.}
  \bibinfo{year}{2002}\natexlab{}.
\newblock \bibinfo{booktitle}{\emph{Technophobia: The psychological impact of
  information technology}}.
\newblock \bibinfo{publisher}{Routledge}.
\newblock


\bibitem[Cheon et~al\mbox{.}(2021)]%
        {cheon2021human}
\bibfield{author}{\bibinfo{person}{EunJeong Cheon}, \bibinfo{person}{Cristina
  Zaga}, \bibinfo{person}{Hee~Rin Lee}, \bibinfo{person}{Maria~Luce Lupetti},
  \bibinfo{person}{Lynn Dombrowski}, {and} \bibinfo{person}{Malte~F Jung}.}
  \bibinfo{year}{2021}\natexlab{}.
\newblock \showarticletitle{Human-Machine Partnerships in the Future of Work:
  Exploring the Role of Emerging Technologies in Future Workplaces}. In
  \bibinfo{booktitle}{\emph{Companion Publication of the 2021 Conference on
  Computer Supported Cooperative Work and Social Computing}}.
  \bibinfo{pages}{323--326}.
\newblock


\bibitem[Clemmensen and Clemmensen(2021)]%
        {clemmensen2021socio}
\bibfield{author}{\bibinfo{person}{Torkil Clemmensen} {and}
  \bibinfo{person}{Torkil Clemmensen}.} \bibinfo{year}{2021}\natexlab{}.
\newblock \showarticletitle{Socio-Technical HCI Design in a Wider Context}.
\newblock \bibinfo{journal}{\emph{Human Work Interaction Design: A Platform for
  Theory and Action}} (\bibinfo{year}{2021}), \bibinfo{pages}{267--280}.
\newblock


\bibitem[Coeckelbergh(2020)]%
        {coeckelbergh2020artificial}
\bibfield{author}{\bibinfo{person}{Mark Coeckelbergh}.}
  \bibinfo{year}{2020}\natexlab{}.
\newblock \showarticletitle{Artificial intelligence, responsibility
  attribution, and a relational justification of explainability}.
\newblock \bibinfo{journal}{\emph{Science and engineering ethics}}
  \bibinfo{volume}{26}, \bibinfo{number}{4} (\bibinfo{year}{2020}),
  \bibinfo{pages}{2051--2068}.
\newblock


\bibitem[Colley et~al\mbox{.}(2020)]%
        {Colley2020EffectOV}
\bibfield{author}{\bibinfo{person}{Mark Colley}, \bibinfo{person}{Christian
  Br{\"a}uner}, \bibinfo{person}{Mirjam Lanzer}, \bibinfo{person}{Marcel
  Walch}, \bibinfo{person}{Martin R.~K. Baumann}, {and} \bibinfo{person}{Enrico
  Rukzio}.} \bibinfo{year}{2020}\natexlab{}.
\newblock \showarticletitle{Effect of Visualization of Pedestrian Intention
  Recognition on Trust and Cognitive Load}.
\newblock \bibinfo{journal}{\emph{12th International Conference on Automotive
  User Interfaces and Interactive Vehicular Applications}}
  (\bibinfo{year}{2020}).
\newblock


\bibitem[Committee et~al\mbox{.}(2014)]%
        {sae2014taxonomy}
\bibfield{author}{\bibinfo{person}{SAE On-Road Automated Vehicle~Standards
  Committee} {et~al\mbox{.}}} \bibinfo{year}{2014}\natexlab{}.
\newblock \showarticletitle{Taxonomy and definitions for terms related to
  on-road motor vehicle automated driving systems}.
\newblock \bibinfo{journal}{\emph{SAE Standard J}}  \bibinfo{volume}{3016}
  (\bibinfo{year}{2014}), \bibinfo{pages}{1--16}.
\newblock


\bibitem[Crompton(2021)]%
        {crompton2021decision}
\bibfield{author}{\bibinfo{person}{Laura Crompton}.}
  \bibinfo{year}{2021}\natexlab{}.
\newblock \showarticletitle{The decision-point-dilemma: Yet another problem of
  responsibility in human-AI interaction}.
\newblock \bibinfo{journal}{\emph{Journal of Responsible Technology}}
  \bibinfo{volume}{7} (\bibinfo{year}{2021}), \bibinfo{pages}{100013}.
\newblock


\bibitem[Cunningham and Regan(2015)]%
        {cunningham2015autonomous}
\bibfield{author}{\bibinfo{person}{Mitchell Cunningham} {and}
  \bibinfo{person}{Michael~A Regan}.} \bibinfo{year}{2015}\natexlab{}.
\newblock \showarticletitle{Autonomous vehicles: human factors issues and
  future research}. In \bibinfo{booktitle}{\emph{Proceedings of the 2015
  Australasian Road safety conference}}, Vol.~\bibinfo{volume}{14}.
\newblock


\bibitem[Dafoe et~al\mbox{.}(2021)]%
        {dafoe2021cooperative}
\bibfield{author}{\bibinfo{person}{Allan Dafoe}, \bibinfo{person}{Yoram
  Bachrach}, \bibinfo{person}{Gillian Hadfield}, \bibinfo{person}{Eric
  Horvitz}, \bibinfo{person}{Kate Larson}, {and} \bibinfo{person}{Thore
  Graepel}.} \bibinfo{year}{2021}\natexlab{}.
\newblock \bibinfo{title}{Cooperative AI: machines must learn to find common
  ground}.
\newblock
\newblock


\bibitem[Demir et~al\mbox{.}(2018)]%
        {demir2018team}
\bibfield{author}{\bibinfo{person}{Mustafa Demir}, \bibinfo{person}{Aaron~D
  Likens}, \bibinfo{person}{Nancy~J Cooke}, \bibinfo{person}{Polemnia~G
  Amazeen}, {and} \bibinfo{person}{Nathan~J McNeese}.}
  \bibinfo{year}{2018}\natexlab{}.
\newblock \showarticletitle{Team coordination and effectiveness in
  human-autonomy teaming}.
\newblock \bibinfo{journal}{\emph{IEEE Transactions on Human-Machine Systems}}
  \bibinfo{volume}{49}, \bibinfo{number}{2} (\bibinfo{year}{2018}),
  \bibinfo{pages}{150--159}.
\newblock


\bibitem[Deo and Trivedi(2019)]%
        {deo2019looking}
\bibfield{author}{\bibinfo{person}{Nachiket Deo} {and} \bibinfo{person}{Mohan~M
  Trivedi}.} \bibinfo{year}{2019}\natexlab{}.
\newblock \showarticletitle{Looking at the driver/rider in autonomous vehicles
  to predict take-over readiness}.
\newblock \bibinfo{journal}{\emph{IEEE Transactions on Intelligent Vehicles}}
  \bibinfo{volume}{5}, \bibinfo{number}{1} (\bibinfo{year}{2019}),
  \bibinfo{pages}{41--52}.
\newblock


\bibitem[Deo and Trivedi(2020)]%
        {Deo2020LookingAT}
\bibfield{author}{\bibinfo{person}{Nachiket Deo} {and}
  \bibinfo{person}{Mohan~Manubhai Trivedi}.} \bibinfo{year}{2020}\natexlab{}.
\newblock \showarticletitle{Looking at the Driver/Rider in Autonomous Vehicles
  to Predict Take-Over Readiness}.
\newblock \bibinfo{journal}{\emph{IEEE Transactions on Intelligent Vehicles}}
  \bibinfo{volume}{5} (\bibinfo{year}{2020}), \bibinfo{pages}{41--52}.
\newblock


\bibitem[Dirsehan and Can(2020)]%
        {Dirsehan2020ExaminationOT}
\bibfield{author}{\bibinfo{person}{Taşkın Dirsehan} {and}
  \bibinfo{person}{Ceren Can}.} \bibinfo{year}{2020}\natexlab{}.
\newblock \showarticletitle{Examination of trust and sustainability concerns in
  autonomous vehicle adoption}.
\newblock \bibinfo{journal}{\emph{Technology in Society}}  \bibinfo{volume}{63}
  (\bibinfo{year}{2020}), \bibinfo{pages}{101361}.
\newblock


\bibitem[Dixit et~al\mbox{.}(2016)]%
        {Dixit2016AutonomousVD}
\bibfield{author}{\bibinfo{person}{Vinayak~V. Dixit}, \bibinfo{person}{Sai
  Chand}, {and} \bibinfo{person}{Divya~Jayakumar Nair}.}
  \bibinfo{year}{2016}\natexlab{}.
\newblock \showarticletitle{Autonomous Vehicles: Disengagements, Accidents and
  Reaction Times}.
\newblock \bibinfo{journal}{\emph{PLoS ONE}}  \bibinfo{volume}{11}
  (\bibinfo{year}{2016}).
\newblock


\bibitem[Downey(2021)]%
        {downey2021partial}
\bibfield{author}{\bibinfo{person}{Mitch Downey}.}
  \bibinfo{year}{2021}\natexlab{}.
\newblock \showarticletitle{Partial automation and the technology-enabled
  deskilling of routine jobs}.
\newblock \bibinfo{journal}{\emph{Labour Economics}}  \bibinfo{volume}{69}
  (\bibinfo{year}{2021}), \bibinfo{pages}{101973}.
\newblock


\bibitem[Drexler et~al\mbox{.}(2019)]%
        {drexler2019handover}
\bibfield{author}{\bibinfo{person}{D{\'a}niel~A Drexler},
  \bibinfo{person}{Arp{\'a}d Tak{\'a}cs}, \bibinfo{person}{Tam{\'a}s~D Nagy},
  {and} \bibinfo{person}{Tam{\'a}s Haidegger}.}
  \bibinfo{year}{2019}\natexlab{}.
\newblock \showarticletitle{Handover Process of Autonomous Vehicles-technology
  and application challenges}.
\newblock \bibinfo{journal}{\emph{Acta Polytechnica Hungarica}}
  \bibinfo{volume}{16}, \bibinfo{number}{9} (\bibinfo{year}{2019}),
  \bibinfo{pages}{235--255}.
\newblock


\bibitem[Du et~al\mbox{.}(2018)]%
        {Du2018VoiceUI}
\bibfield{author}{\bibinfo{person}{Yuemeng Du}, \bibinfo{person}{Jingyan Qin},
  \bibinfo{person}{Shujing Zhang}, \bibinfo{person}{Sha Cao}, {and}
  \bibinfo{person}{Jinhua Dou}.} \bibinfo{year}{2018}\natexlab{}.
\newblock \showarticletitle{Voice User Interface Interaction Design Research
  Based on User Mental Model in Autonomous Vehicle}. In
  \bibinfo{booktitle}{\emph{HCI}}.
\newblock


\bibitem[Duhigg(2016)]%
        {duhigg2016google}
\bibfield{author}{\bibinfo{person}{Charles Duhigg}.}
  \bibinfo{year}{2016}\natexlab{}.
\newblock \showarticletitle{What Google learned from its quest to build the
  perfect team}.
\newblock \bibinfo{journal}{\emph{The New York Times Magazine}}
  \bibinfo{volume}{26}, \bibinfo{number}{2016} (\bibinfo{year}{2016}),
  \bibinfo{pages}{2016}.
\newblock


\bibitem[Ebnali et~al\mbox{.}(2019)]%
        {ebnali2019user}
\bibfield{author}{\bibinfo{person}{Mahdi Ebnali}, \bibinfo{person}{Cyrus Kian},
  \bibinfo{person}{Majid Ebnali-Heidari}, {and} \bibinfo{person}{Adel
  Mazloumi}.} \bibinfo{year}{2019}\natexlab{}.
\newblock \showarticletitle{User experience in immersive VR-based serious game:
  an application in highly automated driving training}. In
  \bibinfo{booktitle}{\emph{International Conference on Applied Human Factors
  and Ergonomics}}. Springer, \bibinfo{pages}{133--144}.
\newblock


\bibitem[Ekman et~al\mbox{.}(2016)]%
        {Ekman2016CreatingAT}
\bibfield{author}{\bibinfo{person}{Fredrick Ekman}, \bibinfo{person}{Mikael
  Johansson}, {and} \bibinfo{person}{Jana Sochor}.}
  \bibinfo{year}{2016}\natexlab{}.
\newblock \showarticletitle{Creating Appropriate Trust for Autonomous Vehicle
  Systems: A Framework for HMI Design}.
\newblock


\bibitem[Ekwoge et~al\mbox{.}(2017)]%
        {ekwoge2017tester}
\bibfield{author}{\bibinfo{person}{Oswald~Mesumbe Ekwoge},
  \bibinfo{person}{Awdren Font{\~a}o}, {and} \bibinfo{person}{Arilo~C
  Dias-Neto}.} \bibinfo{year}{2017}\natexlab{}.
\newblock \showarticletitle{Tester experience: concept, issues and definition}.
  In \bibinfo{booktitle}{\emph{2017 ieee 41st annual computer software and
  applications conference (compsac)}}, Vol.~\bibinfo{volume}{1}. IEEE,
  \bibinfo{pages}{208--213}.
\newblock


\bibitem[Gibson et~al\mbox{.}(2016)]%
        {gibson2016situation}
\bibfield{author}{\bibinfo{person}{Madeleine Gibson}, \bibinfo{person}{John
  Lee}, \bibinfo{person}{Vindhya Venkatraman}, \bibinfo{person}{Morgan Price},
  \bibinfo{person}{Jeffrey Lewis}, \bibinfo{person}{Olivia Montgomery},
  \bibinfo{person}{Bilge Mutlu}, \bibinfo{person}{Joshua Domeyer}, {and}
  \bibinfo{person}{James Foley}.} \bibinfo{year}{2016}\natexlab{}.
\newblock \showarticletitle{Situation awareness, scenarios, and secondary
  tasks: measuring driver performance and safety margins in highly automated
  vehicles}.
\newblock \bibinfo{journal}{\emph{SAE International Journal of Passenger
  Cars-Electronic and Electrical Systems}} \bibinfo{volume}{9},
  \bibinfo{number}{1} (\bibinfo{year}{2016}), \bibinfo{pages}{237--243}.
\newblock


\bibitem[Gyevnar et~al\mbox{.}(2022)]%
        {Gyevnar2022AHM}
\bibfield{author}{\bibinfo{person}{Balint Gyevnar},
  \bibinfo{person}{Massimiliano Tamborski}, \bibinfo{person}{Cheng-Hsien Wang},
  \bibinfo{person}{Christopher~G. Lucas}, \bibinfo{person}{Shay~B. Cohen},
  {and} \bibinfo{person}{Stefano~V. Albrecht}.}
  \bibinfo{year}{2022}\natexlab{}.
\newblock \showarticletitle{A Human-Centric Method for Generating Causal
  Explanations in Natural Language for Autonomous Vehicle Motion Planning}.
\newblock \bibinfo{journal}{\emph{ArXiv}}  \bibinfo{volume}{abs/2206.08783}
  (\bibinfo{year}{2022}).
\newblock


\bibitem[Hester et~al\mbox{.}(2017)]%
        {Hester2017DriverTO}
\bibfield{author}{\bibinfo{person}{Michelle Hester}, \bibinfo{person}{Kevin
  Lee}, {and} \bibinfo{person}{Brian~P. Dyre}.}
  \bibinfo{year}{2017}\natexlab{}.
\newblock \showarticletitle{“Driver Take Over”: A Preliminary Exploration
  of Driver Trust and Performance in Autonomous Vehicles}.
\newblock \bibinfo{journal}{\emph{Proceedings of the Human Factors and
  Ergonomics Society Annual Meeting}}  \bibinfo{volume}{61}
  (\bibinfo{year}{2017}), \bibinfo{pages}{1969 -- 1973}.
\newblock


\bibitem[Hewitt et~al\mbox{.}(2019)]%
        {Hewitt2019AssessingPP}
\bibfield{author}{\bibinfo{person}{Charles~P. Hewitt}, \bibinfo{person}{Ioannis
  Politis}, \bibinfo{person}{Theocharis Amanatidis}, {and}
  \bibinfo{person}{Advait Sarkar}.} \bibinfo{year}{2019}\natexlab{}.
\newblock \showarticletitle{Assessing public perception of self-driving cars:
  the autonomous vehicle acceptance model}.
\newblock \bibinfo{journal}{\emph{Proceedings of the 24th International
  Conference on Intelligent User Interfaces}} (\bibinfo{year}{2019}).
\newblock


\bibitem[Huang and Pitts(2020)]%
        {huang2020age}
\bibfield{author}{\bibinfo{person}{Gaojian Huang} {and}
  \bibinfo{person}{Brandon Pitts}.} \bibinfo{year}{2020}\natexlab{}.
\newblock \showarticletitle{Age-related differences in takeover request
  modality preferences and attention allocation during semi-autonomous
  driving}. In \bibinfo{booktitle}{\emph{International Conference on
  Human-Computer Interaction}}. Springer, \bibinfo{pages}{135--146}.
\newblock


\bibitem[Huesemann and Huesemann(2011)]%
        {huesemann2011techno}
\bibfield{author}{\bibinfo{person}{Michael Huesemann} {and}
  \bibinfo{person}{Joyce Huesemann}.} \bibinfo{year}{2011}\natexlab{}.
\newblock \bibinfo{booktitle}{\emph{Techno-fix: why technology won't save us or
  the environment}}.
\newblock \bibinfo{publisher}{New Society Publishers}.
\newblock


\bibitem[Kalra and Paddock(2016)]%
        {Kalra2016DrivingTS}
\bibfield{author}{\bibinfo{person}{Nidhi Kalra} {and} \bibinfo{person}{Susan~M.
  Paddock}.} \bibinfo{year}{2016}\natexlab{}.
\newblock \showarticletitle{Driving to safety: How many miles of driving would
  it take to demonstrate autonomous vehicle reliability?}
\newblock \bibinfo{journal}{\emph{Transportation Research Part A-policy and
  Practice}}  \bibinfo{volume}{94} (\bibinfo{year}{2016}),
  \bibinfo{pages}{182--193}.
\newblock


\bibitem[Karvonen et~al\mbox{.}(2011)]%
        {karvonen2011hidden}
\bibfield{author}{\bibinfo{person}{Hannu Karvonen}, \bibinfo{person}{Iina
  Aaltonen}, \bibinfo{person}{Mikael Wahlstr{\"o}m}, \bibinfo{person}{Leena
  Salo}, \bibinfo{person}{Paula Savioja}, {and} \bibinfo{person}{Leena
  Norros}.} \bibinfo{year}{2011}\natexlab{}.
\newblock \showarticletitle{Hidden roles of the train driver: A challenge for
  metro automation}.
\newblock \bibinfo{journal}{\emph{Interacting with computers}}
  \bibinfo{volume}{23}, \bibinfo{number}{4} (\bibinfo{year}{2011}),
  \bibinfo{pages}{289--298}.
\newblock


\bibitem[Kim and Yang(2017)]%
        {kim2017takeover}
\bibfield{author}{\bibinfo{person}{Hyung~Jun Kim} {and}
  \bibinfo{person}{Ji~Hyun Yang}.} \bibinfo{year}{2017}\natexlab{}.
\newblock \showarticletitle{Takeover requests in simulated partially autonomous
  vehicles considering human factors}.
\newblock \bibinfo{journal}{\emph{IEEE Transactions on Human-Machine Systems}}
  \bibinfo{volume}{47}, \bibinfo{number}{5} (\bibinfo{year}{2017}),
  \bibinfo{pages}{735--740}.
\newblock


\bibitem[Kim et~al\mbox{.}(2018)]%
        {Kim2018TakeoverPA}
\bibfield{author}{\bibinfo{person}{Jungsook Kim}, \bibinfo{person}{Hyunsuk
  Kim}, \bibinfo{person}{Woojin Kim}, {and} \bibinfo{person}{Daesub Yoon}.}
  \bibinfo{year}{2018}\natexlab{}.
\newblock \showarticletitle{Take-over performance analysis depending on the
  drivers’ non-driving secondary tasks in automated vehicles}.
\newblock \bibinfo{journal}{\emph{2018 International Conference on Information
  and Communication Technology Convergence (ICTC)}} (\bibinfo{year}{2018}),
  \bibinfo{pages}{1364--1366}.
\newblock


\bibitem[Kishore~Bhoopalam et~al\mbox{.}(2021)]%
        {kishore2021long}
\bibfield{author}{\bibinfo{person}{Anirudh Kishore~Bhoopalam},
  \bibinfo{person}{Roy van~den Berg}, \bibinfo{person}{Niels Agatz}, {and}
  \bibinfo{person}{Caspar Chorus}.} \bibinfo{year}{2021}\natexlab{}.
\newblock \showarticletitle{The long road to automated trucking: Insights from
  driver focus groups}.
\newblock \bibinfo{journal}{\emph{Roy and Agatz, Niels AH and Chorus, Caspar,
  The Long Road to Automated Trucking: Insights from Driver Focus Groups
  (February 4, 2021)}} (\bibinfo{year}{2021}).
\newblock


\bibitem[Koo et~al\mbox{.}(2015)]%
        {Koo2015WhyDM}
\bibfield{author}{\bibinfo{person}{Jeamin Koo}, \bibinfo{person}{Jungsuk Kwac},
  \bibinfo{person}{Wendy Ju}, \bibinfo{person}{Martin Steinert},
  \bibinfo{person}{Larry~J. Leifer}, {and} \bibinfo{person}{Clifford Nass}.}
  \bibinfo{year}{2015}\natexlab{}.
\newblock \showarticletitle{Why did my car just do that? Explaining
  semi-autonomous driving actions to improve driver understanding, trust, and
  performance}.
\newblock \bibinfo{journal}{\emph{International Journal on Interactive Design
  and Manufacturing (IJIDeM)}}  \bibinfo{volume}{9} (\bibinfo{year}{2015}),
  \bibinfo{pages}{269--275}.
\newblock


\bibitem[Lee et~al\mbox{.}(2017)]%
        {lee2017study}
\bibfield{author}{\bibinfo{person}{Jiin-in Lee}, \bibinfo{person}{Na-eun Kim},
  {and} \bibinfo{person}{Jin-woo Kim}.} \bibinfo{year}{2017}\natexlab{}.
\newblock \showarticletitle{A study on driver experience in autonomous car
  based on trust and distrust model of automation system}.
\newblock \bibinfo{journal}{\emph{Journal of Digital Contents Society}}
  \bibinfo{volume}{18}, \bibinfo{number}{4} (\bibinfo{year}{2017}),
  \bibinfo{pages}{713--722}.
\newblock


\bibitem[Li et~al\mbox{.}(2018)]%
        {Li2018SwitchedCD}
\bibfield{author}{\bibinfo{person}{Y. Li}, \bibinfo{person}{Dihua Sun},
  \bibinfo{person}{Min Zhao}, \bibinfo{person}{Dong Chen},
  \bibinfo{person}{Senlin Cheng}, {and} \bibinfo{person}{Fei Xie}.}
  \bibinfo{year}{2018}\natexlab{}.
\newblock \showarticletitle{Switched Cooperative Driving Model towards Human
  Vehicle Copiloting Situation: A Cyberphysical Perspective}.
\newblock \bibinfo{journal}{\emph{Journal of Advanced Transportation}}
  \bibinfo{volume}{2018} (\bibinfo{year}{2018}), \bibinfo{pages}{1--11}.
\newblock


\bibitem[Liang et~al\mbox{.}(2019)]%
        {liang2019implicit}
\bibfield{author}{\bibinfo{person}{Claire Liang}, \bibinfo{person}{Julia
  Proft}, \bibinfo{person}{Erik Andersen}, {and} \bibinfo{person}{Ross~A
  Knepper}.} \bibinfo{year}{2019}\natexlab{}.
\newblock \showarticletitle{Implicit communication of actionable information in
  human-ai teams}. In \bibinfo{booktitle}{\emph{Proceedings of the 2019 CHI
  Conference on Human Factors in Computing Systems}}. \bibinfo{pages}{1--13}.
\newblock


\bibitem[Lindemann et~al\mbox{.}(2018)]%
        {Lindemann2018CatchMD}
\bibfield{author}{\bibinfo{person}{Patrick Lindemann},
  \bibinfo{person}{Tae-Young Lee}, {and} \bibinfo{person}{Gerhard Rigoll}.}
  \bibinfo{year}{2018}\natexlab{}.
\newblock \showarticletitle{Catch My Drift: Elevating Situation Awareness for
  Highly Automated Driving with an Explanatory Windshield Display User
  Interface}.
\newblock \bibinfo{journal}{\emph{Multimodal Technol. Interact.}}
  \bibinfo{volume}{2} (\bibinfo{year}{2018}), \bibinfo{pages}{71}.
\newblock


\bibitem[Llaneras et~al\mbox{.}(2017)]%
        {llaneras2017strategies}
\bibfield{author}{\bibinfo{person}{Robert~E Llaneras}, \bibinfo{person}{Brad~R
  Cannon}, {and} \bibinfo{person}{Charles~A Green}.}
  \bibinfo{year}{2017}\natexlab{}.
\newblock \showarticletitle{Strategies to assist drivers in remaining attentive
  while under partially automated driving: Verification of human--machine
  interface concepts}.
\newblock \bibinfo{journal}{\emph{Transportation research record}}
  \bibinfo{volume}{2663}, \bibinfo{number}{1} (\bibinfo{year}{2017}),
  \bibinfo{pages}{20--26}.
\newblock


\bibitem[Long and Magerko(2020)]%
        {long2020ai}
\bibfield{author}{\bibinfo{person}{Duri Long} {and} \bibinfo{person}{Brian
  Magerko}.} \bibinfo{year}{2020}\natexlab{}.
\newblock \showarticletitle{What is AI literacy? Competencies and design
  considerations}. In \bibinfo{booktitle}{\emph{Proceedings of the 2020 CHI
  conference on human factors in computing systems}}. \bibinfo{pages}{1--16}.
\newblock


\bibitem[Lyons(2013)]%
        {Lyons2013BeingTA}
\bibfield{author}{\bibinfo{person}{Joseph~B. Lyons}.}
  \bibinfo{year}{2013}\natexlab{}.
\newblock \showarticletitle{Being Transparent about Transparency: A Model for
  Human-Robot Interaction}. In \bibinfo{booktitle}{\emph{AAAI Spring Symposium:
  Trust and Autonomous Systems}}.
\newblock


\bibitem[Magnisalis et~al\mbox{.}(2011)]%
        {magnisalis2011adaptive}
\bibfield{author}{\bibinfo{person}{Ioannis Magnisalis},
  \bibinfo{person}{Stavros Demetriadis}, {and} \bibinfo{person}{Anastasios
  Karakostas}.} \bibinfo{year}{2011}\natexlab{}.
\newblock \showarticletitle{Adaptive and intelligent systems for collaborative
  learning support: A review of the field}.
\newblock \bibinfo{journal}{\emph{IEEE transactions on Learning Technologies}}
  \bibinfo{volume}{4}, \bibinfo{number}{1} (\bibinfo{year}{2011}),
  \bibinfo{pages}{5--20}.
\newblock


\bibitem[Maguire and Delahunt(2017)]%
        {maguire2017doing}
\bibfield{author}{\bibinfo{person}{Moira Maguire} {and} \bibinfo{person}{Brid
  Delahunt}.} \bibinfo{year}{2017}\natexlab{}.
\newblock \showarticletitle{Doing a thematic analysis: A practical,
  step-by-step guide for learning and teaching scholars.}
\newblock \bibinfo{journal}{\emph{All Ireland Journal of Higher Education}}
  \bibinfo{volume}{9}, \bibinfo{number}{3} (\bibinfo{year}{2017}).
\newblock


\bibitem[Makris et~al\mbox{.}(2016)]%
        {makris2016augmented}
\bibfield{author}{\bibinfo{person}{Sotiris Makris}, \bibinfo{person}{Panagiotis
  Karagiannis}, \bibinfo{person}{Spyridon Koukas}, {and}
  \bibinfo{person}{Aleksandros-Stereos Matthaiakis}.}
  \bibinfo{year}{2016}\natexlab{}.
\newblock \showarticletitle{Augmented reality system for operator support in
  human--robot collaborative assembly}.
\newblock \bibinfo{journal}{\emph{CIRP Annals}} \bibinfo{volume}{65},
  \bibinfo{number}{1} (\bibinfo{year}{2016}), \bibinfo{pages}{61--64}.
\newblock


\bibitem[Manyika et~al\mbox{.}(2017)]%
        {manyika2017jobs}
\bibfield{author}{\bibinfo{person}{James Manyika}, \bibinfo{person}{Susan
  Lund}, \bibinfo{person}{Michael Chui}, \bibinfo{person}{Jacques Bughin},
  \bibinfo{person}{Jonathan Woetzel}, \bibinfo{person}{Parul Batra},
  \bibinfo{person}{Ryan Ko}, {and} \bibinfo{person}{Saurabh Sanghvi}.}
  \bibinfo{year}{2017}\natexlab{}.
\newblock \showarticletitle{Jobs lost, jobs gained: Workforce transitions in a
  time of automation}.
\newblock \bibinfo{journal}{\emph{McKinsey Global Institute}}
  \bibinfo{volume}{150} (\bibinfo{year}{2017}).
\newblock


\bibitem[McDonald et~al\mbox{.}(2019)]%
        {mcdonald2019reliability}
\bibfield{author}{\bibinfo{person}{Nora McDonald}, \bibinfo{person}{Sarita
  Schoenebeck}, {and} \bibinfo{person}{Andrea Forte}.}
  \bibinfo{year}{2019}\natexlab{}.
\newblock \showarticletitle{Reliability and inter-rater reliability in
  qualitative research: Norms and guidelines for CSCW and HCI practice}.
\newblock \bibinfo{journal}{\emph{Proceedings of the ACM on human-computer
  interaction}} \bibinfo{volume}{3}, \bibinfo{number}{CSCW}
  (\bibinfo{year}{2019}), \bibinfo{pages}{1--23}.
\newblock


\bibitem[McGehee et~al\mbox{.}(2016)]%
        {mcgehee2016review}
\bibfield{author}{\bibinfo{person}{Daniel~V McGehee}, \bibinfo{person}{Mark
  Brewer}, \bibinfo{person}{Chris Schwarz}, \bibinfo{person}{Bryant~Walker
  Smith}, {et~al\mbox{.}}} \bibinfo{year}{2016}\natexlab{}.
\newblock \bibinfo{booktitle}{\emph{Review of automated vehicle technology:
  policy and implementation implications.}}
\newblock \bibinfo{type}{{T}echnical {R}eport}. \bibinfo{institution}{Iowa.
  Dept. of Transportation}.
\newblock


\bibitem[McGuinness et~al\mbox{.}(2008)]%
        {mcguinness2008characteristics}
\bibfield{author}{\bibinfo{person}{Seamus McGuinness},
  \bibinfo{person}{John~Will Freebairn}, {and} \bibinfo{person}{Kostas~G
  Mavromaras}.} \bibinfo{year}{2008}\natexlab{}.
\newblock \bibinfo{booktitle}{\emph{Characteristics of minimum wage
  employees}}.
\newblock \bibinfo{publisher}{Australian Fair Pay Commission}.
\newblock


\bibitem[Meguia et~al\mbox{.}(2019)]%
        {Meguia2019PrinciplesOT}
\bibfield{author}{\bibinfo{person}{Raissa~Pokam Meguia}, \bibinfo{person}{Serge
  Debernard}, \bibinfo{person}{Christine Chauvin}, {and}
  \bibinfo{person}{Sabine Langlois}.} \bibinfo{year}{2019}\natexlab{}.
\newblock \showarticletitle{Principles of transparency for autonomous vehicles:
  first results of an experiment with an augmented reality human–machine
  interface}.
\newblock \bibinfo{journal}{\emph{Cognition, Technology \& Work}}
  (\bibinfo{year}{2019}), \bibinfo{pages}{1--14}.
\newblock


\bibitem[Millard‐Ball(2016)]%
        {MillardBall2016PedestriansAV}
\bibfield{author}{\bibinfo{person}{Adam Millard‐Ball}.}
  \bibinfo{year}{2016}\natexlab{}.
\newblock \showarticletitle{Pedestrians, Autonomous Vehicles, and Cities}.
\newblock \bibinfo{journal}{\emph{Journal of Planning Education and Research}}
  \bibinfo{volume}{38} (\bibinfo{year}{2016}), \bibinfo{pages}{12 -- 6}.
\newblock


\bibitem[Mirnig et~al\mbox{.}(2016)]%
        {mirnig2016framework}
\bibfield{author}{\bibinfo{person}{Alexander~G Mirnig},
  \bibinfo{person}{Philipp Wintersberger}, \bibinfo{person}{Christine Sutter},
  {and} \bibinfo{person}{J{\"u}rgen Ziegler}.} \bibinfo{year}{2016}\natexlab{}.
\newblock \showarticletitle{A framework for analyzing and calibrating trust in
  automated vehicles}. In \bibinfo{booktitle}{\emph{Adjunct proceedings of the
  8th international conference on automotive user interfaces and interactive
  vehicular applications}}. \bibinfo{pages}{33--38}.
\newblock


\bibitem[Mok et~al\mbox{.}(2015)]%
        {mok2015emergency}
\bibfield{author}{\bibinfo{person}{Brian Mok}, \bibinfo{person}{Mishel Johns},
  \bibinfo{person}{Key~Jung Lee}, \bibinfo{person}{David Miller},
  \bibinfo{person}{David Sirkin}, \bibinfo{person}{Page Ive}, {and}
  \bibinfo{person}{Wendy Ju}.} \bibinfo{year}{2015}\natexlab{}.
\newblock \showarticletitle{Emergency, automation off: Unstructured transition
  timing for distracted drivers of automated vehicles}. In
  \bibinfo{booktitle}{\emph{2015 IEEE 18th international conference on
  intelligent transportation systems}}. IEEE, \bibinfo{pages}{2458--2464}.
\newblock


\bibitem[Moniz and Krings(2014)]%
        {moniz2014technology}
\bibfield{author}{\bibinfo{person}{Ant{\'o}nio Moniz} {and}
  \bibinfo{person}{Bettina-Johanna Krings}.} \bibinfo{year}{2014}\natexlab{}.
\newblock \showarticletitle{Technology assessment approach to human-robot
  interactions in work environments}. In \bibinfo{booktitle}{\emph{2014 7th
  International Conference on Human System Interactions (HSI)}}. IEEE,
  \bibinfo{pages}{282--289}.
\newblock


\bibitem[Mueller et~al\mbox{.}(2019)]%
        {mueller2019explanation}
\bibfield{author}{\bibinfo{person}{Shane~T Mueller}, \bibinfo{person}{Robert~R
  Hoffman}, \bibinfo{person}{William Clancey}, \bibinfo{person}{Abigail Emrey},
  {and} \bibinfo{person}{Gary Klein}.} \bibinfo{year}{2019}\natexlab{}.
\newblock \showarticletitle{Explanation in human-AI systems: A literature
  meta-review, synopsis of key ideas and publications, and bibliography for
  explainable AI}.
\newblock \bibinfo{journal}{\emph{arXiv preprint arXiv:1902.01876}}
  (\bibinfo{year}{2019}).
\newblock


\bibitem[Muller et~al\mbox{.}(2021)]%
        {muller2021designing}
\bibfield{author}{\bibinfo{person}{Michael Muller},
  \bibinfo{person}{Christine~T Wolf}, \bibinfo{person}{Josh Andres},
  \bibinfo{person}{Michael Desmond}, \bibinfo{person}{Narendra~Nath Joshi},
  \bibinfo{person}{Zahra Ashktorab}, \bibinfo{person}{Aabhas Sharma},
  \bibinfo{person}{Kristina Brimijoin}, \bibinfo{person}{Qian Pan},
  \bibinfo{person}{Evelyn Duesterwald}, {et~al\mbox{.}}}
  \bibinfo{year}{2021}\natexlab{}.
\newblock \showarticletitle{Designing ground truth and the social life of
  labels}. In \bibinfo{booktitle}{\emph{Proceedings of the 2021 CHI Conference
  on Human Factors in Computing Systems}}. \bibinfo{pages}{1--16}.
\newblock


\bibitem[Orii et~al\mbox{.}(2021)]%
        {orii2021perceptions}
\bibfield{author}{\bibinfo{person}{Lisa Orii}, \bibinfo{person}{Diana Tosca},
  \bibinfo{person}{Andrew~L Kun}, {and} \bibinfo{person}{Orit Shaer}.}
  \bibinfo{year}{2021}\natexlab{}.
\newblock \showarticletitle{Perceptions on the Future of Automation in
  r/Truckers}. In \bibinfo{booktitle}{\emph{Extended abstracts of the 2021 CHI
  conference on human factors in computing systems}}. \bibinfo{pages}{1--6}.
\newblock


\bibitem[Ososky et~al\mbox{.}(2012)]%
        {ososky2012importance}
\bibfield{author}{\bibinfo{person}{Scott Ososky}, \bibinfo{person}{David
  Schuster}, \bibinfo{person}{Florian Jentsch}, \bibinfo{person}{Stephen
  Fiore}, \bibinfo{person}{Randall Shumaker}, \bibinfo{person}{Christian
  Lebiere}, \bibinfo{person}{Unmesh Kurup}, \bibinfo{person}{Jean Oh}, {and}
  \bibinfo{person}{Anthony Stentz}.} \bibinfo{year}{2012}\natexlab{}.
\newblock \showarticletitle{The importance of shared mental models and shared
  situation awareness for transforming robots from tools to teammates}. In
  \bibinfo{booktitle}{\emph{Unmanned systems technology XIV}},
  Vol.~\bibinfo{volume}{8387}. SPIE, \bibinfo{pages}{397--408}.
\newblock


\bibitem[Palinkas et~al\mbox{.}(2015)]%
        {palinkas2015purposeful}
\bibfield{author}{\bibinfo{person}{Lawrence~A Palinkas},
  \bibinfo{person}{Sarah~M Horwitz}, \bibinfo{person}{Carla~A Green},
  \bibinfo{person}{Jennifer~P Wisdom}, \bibinfo{person}{Naihua Duan}, {and}
  \bibinfo{person}{Kimberly Hoagwood}.} \bibinfo{year}{2015}\natexlab{}.
\newblock \showarticletitle{Purposeful sampling for qualitative data collection
  and analysis in mixed method implementation research}.
\newblock \bibinfo{journal}{\emph{Administration and policy in mental health
  and mental health services research}} \bibinfo{volume}{42},
  \bibinfo{number}{5} (\bibinfo{year}{2015}), \bibinfo{pages}{533--544}.
\newblock


\bibitem[Panagiotopoulos et~al\mbox{.}(2020)]%
        {Panagiotopoulos2020AreCR}
\bibfield{author}{\bibinfo{person}{Ilias~E. Panagiotopoulos},
  \bibinfo{person}{George~J. Dimitrakopoulos}, \bibinfo{person}{Gabriele
  Keraite}, {and} \bibinfo{person}{Urte Steikuniene}.}
  \bibinfo{year}{2020}\natexlab{}.
\newblock \showarticletitle{Are Consumers Ready to Adopt Highly Automated
  Passenger Vehicles? Results from a Cross-national Survey in Europe}. In
  \bibinfo{booktitle}{\emph{VEHITS}}.
\newblock


\bibitem[Ramchurn et~al\mbox{.}(2021)]%
        {ramchurn2021trustworthy}
\bibfield{author}{\bibinfo{person}{Sarvapali~D Ramchurn},
  \bibinfo{person}{Sebastian Stein}, {and} \bibinfo{person}{Nicholas~R
  Jennings}.} \bibinfo{year}{2021}\natexlab{}.
\newblock \showarticletitle{Trustworthy human-AI partnerships}.
\newblock \bibinfo{journal}{\emph{Iscience}} \bibinfo{volume}{24},
  \bibinfo{number}{8} (\bibinfo{year}{2021}), \bibinfo{pages}{102891}.
\newblock


\bibitem[Ribera and Lapedriza(2019)]%
        {ribera2019can}
\bibfield{author}{\bibinfo{person}{Mireia Ribera} {and} \bibinfo{person}{Agata
  Lapedriza}.} \bibinfo{year}{2019}\natexlab{}.
\newblock \showarticletitle{Can we do better explanations? A proposal of
  user-centered explainable AI.}. In \bibinfo{booktitle}{\emph{IUI Workshops}},
  Vol.~\bibinfo{volume}{2327}. \bibinfo{pages}{38}.
\newblock


\bibitem[Rubin and Rubin(2011)]%
        {rubin2011qualitative}
\bibfield{author}{\bibinfo{person}{Herbert~J Rubin} {and}
  \bibinfo{person}{Irene~S Rubin}.} \bibinfo{year}{2011}\natexlab{}.
\newblock \bibinfo{booktitle}{\emph{Qualitative interviewing: The art of
  hearing data}}.
\newblock \bibinfo{publisher}{sage}.
\newblock


\bibitem[Sambasivan et~al\mbox{.}(2021)]%
        {sambasivan2021everyone}
\bibfield{author}{\bibinfo{person}{Nithya Sambasivan}, \bibinfo{person}{Shivani
  Kapania}, \bibinfo{person}{Hannah Highfill}, \bibinfo{person}{Diana Akrong},
  \bibinfo{person}{Praveen Paritosh}, {and} \bibinfo{person}{Lora~M Aroyo}.}
  \bibinfo{year}{2021}\natexlab{}.
\newblock \showarticletitle{“Everyone wants to do the model work, not the
  data work”: Data Cascades in High-Stakes AI}. In
  \bibinfo{booktitle}{\emph{proceedings of the 2021 CHI Conference on Human
  Factors in Computing Systems}}. \bibinfo{pages}{1--15}.
\newblock


\bibitem[Schelble et~al\mbox{.}(2022)]%
        {schelble2022let}
\bibfield{author}{\bibinfo{person}{Beau~G Schelble},
  \bibinfo{person}{Christopher Flathmann}, \bibinfo{person}{Nathan~J McNeese},
  \bibinfo{person}{Guo Freeman}, {and} \bibinfo{person}{Rohit Mallick}.}
  \bibinfo{year}{2022}\natexlab{}.
\newblock \showarticletitle{Let's Think Together! Assessing Shared Mental
  Models, Performance, and Trust in Human-Agent Teams}.
\newblock \bibinfo{journal}{\emph{Proceedings of the ACM on Human-Computer
  Interaction}} \bibinfo{volume}{6}, \bibinfo{number}{GROUP}
  (\bibinfo{year}{2022}), \bibinfo{pages}{1--29}.
\newblock


\bibitem[Shahrdar et~al\mbox{.}(2018)]%
        {shahrdar2018survey}
\bibfield{author}{\bibinfo{person}{Shervin Shahrdar}, \bibinfo{person}{Luiza
  Menezes}, {and} \bibinfo{person}{Mehrdad Nojoumian}.}
  \bibinfo{year}{2018}\natexlab{}.
\newblock \showarticletitle{A survey on trust in autonomous systems}. In
  \bibinfo{booktitle}{\emph{Science and Information Conference}}. Springer,
  \bibinfo{pages}{368--386}.
\newblock


\bibitem[Shen et~al\mbox{.}(2020)]%
        {Shen2020ToEO}
\bibfield{author}{\bibinfo{person}{Yuan Shen}, \bibinfo{person}{Shanduojiao
  Jiang}, \bibinfo{person}{Yanlin Chen}, \bibinfo{person}{Eileen~Jianxun Yang},
  \bibinfo{person}{Xilun Jin}, \bibinfo{person}{Yuliang Fan}, {and}
  \bibinfo{person}{Katherine~Driggs Campbell}.}
  \bibinfo{year}{2020}\natexlab{}.
\newblock \showarticletitle{To Explain or Not to Explain: A Study on the
  Necessity of Explanations for Autonomous Vehicles}.
\newblock \bibinfo{journal}{\emph{ArXiv}}  \bibinfo{volume}{abs/2006.11684}
  (\bibinfo{year}{2020}).
\newblock


\bibitem[Shively et~al\mbox{.}(2017)]%
        {shively2017human}
\bibfield{author}{\bibinfo{person}{R~Jay Shively}, \bibinfo{person}{Joel
  Lachter}, \bibinfo{person}{Summer~L Brandt}, \bibinfo{person}{Michael
  Matessa}, \bibinfo{person}{Vernol Battiste}, {and} \bibinfo{person}{Walter~W
  Johnson}.} \bibinfo{year}{2017}\natexlab{}.
\newblock \showarticletitle{Why human-autonomy teaming?}. In
  \bibinfo{booktitle}{\emph{International conference on applied human factors
  and ergonomics}}. Springer, \bibinfo{pages}{3--11}.
\newblock


\bibitem[Sonboli et~al\mbox{.}(2021)]%
        {sonboli2021fairness}
\bibfield{author}{\bibinfo{person}{Nasim Sonboli}, \bibinfo{person}{Jessie~J
  Smith}, \bibinfo{person}{Florencia Cabral~Berenfus}, \bibinfo{person}{Robin
  Burke}, {and} \bibinfo{person}{Casey Fiesler}.}
  \bibinfo{year}{2021}\natexlab{}.
\newblock \showarticletitle{Fairness and transparency in recommendation: The
  users’ perspective}. In \bibinfo{booktitle}{\emph{Proceedings of the 29th
  ACM Conference on User Modeling, Adaptation and Personalization}}.
  \bibinfo{pages}{274--279}.
\newblock


\bibitem[Strobl and Thornton(2002)]%
        {strobl2002large}
\bibfield{author}{\bibinfo{person}{Eric Strobl} {and} \bibinfo{person}{Robert~J
  Thornton}.} \bibinfo{year}{2002}\natexlab{}.
\newblock \showarticletitle{Do large employers pay more in developing
  countries? The case of five African countries}.
\newblock \bibinfo{journal}{\emph{The Case of Five African Countries (December
  2002)}} (\bibinfo{year}{2002}).
\newblock


\bibitem[Sunstein(2018)]%
        {Sunstein2018AlgorithmsCB}
\bibfield{author}{\bibinfo{person}{Cass~Robert Sunstein}.}
  \bibinfo{year}{2018}\natexlab{}.
\newblock \showarticletitle{Algorithms, Correcting Biases}.
\newblock \bibinfo{journal}{\emph{Artificial Intelligence - Law}}
  (\bibinfo{year}{2018}).
\newblock


\bibitem[Tak{\'a}cs et~al\mbox{.}(2018)]%
        {takacs2018assessment}
\bibfield{author}{\bibinfo{person}{{\'A}rp{\'a}d Tak{\'a}cs},
  \bibinfo{person}{D{\'a}niel~Andr{\'a}s Drexler}, \bibinfo{person}{P{\'e}ter
  Galambos}, \bibinfo{person}{Imre~J Rudas}, {and} \bibinfo{person}{Tam{\'a}s
  Haidegger}.} \bibinfo{year}{2018}\natexlab{}.
\newblock \showarticletitle{Assessment and standardization of autonomous
  vehicles}. In \bibinfo{booktitle}{\emph{2018 IEEE 22nd International
  Conference on Intelligent Engineering Systems (INES)}}. IEEE,
  \bibinfo{pages}{000185--000192}.
\newblock


\bibitem[Tran et~al\mbox{.}(2018)]%
        {tran2018human}
\bibfield{author}{\bibinfo{person}{Duy Tran}, \bibinfo{person}{Jianhao Du},
  \bibinfo{person}{Weihua Sheng}, \bibinfo{person}{Denis Osipychev},
  \bibinfo{person}{Yuge Sun}, {and} \bibinfo{person}{He Bai}.}
  \bibinfo{year}{2018}\natexlab{}.
\newblock \showarticletitle{A human-vehicle collaborative driving framework for
  driver assistance}.
\newblock \bibinfo{journal}{\emph{IEEE Transactions on Intelligent
  Transportation Systems}} \bibinfo{volume}{20}, \bibinfo{number}{9}
  (\bibinfo{year}{2018}), \bibinfo{pages}{3470--3485}.
\newblock


\bibitem[Trende et~al\mbox{.}(2019)]%
        {Trende2019AnII}
\bibfield{author}{\bibinfo{person}{Alexander Trende}, \bibinfo{person}{Anirudh
  Unni}, \bibinfo{person}{Lars Weber}, \bibinfo{person}{Jochem~W. Rieger},
  {and} \bibinfo{person}{Andreas L{\"u}dtke}.} \bibinfo{year}{2019}\natexlab{}.
\newblock \showarticletitle{An investigation into human-autonomous vs.
  human-human vehicle interaction in time-critical situations}.
\newblock \bibinfo{journal}{\emph{Proceedings of the 12th ACM International
  Conference on PErvasive Technologies Related to Assistive Environments}}
  (\bibinfo{year}{2019}).
\newblock


\bibitem[Tubaro et~al\mbox{.}(2020)]%
        {tubaro2020trainer}
\bibfield{author}{\bibinfo{person}{Paola Tubaro}, \bibinfo{person}{Antonio~A
  Casilli}, {and} \bibinfo{person}{Marion Coville}.}
  \bibinfo{year}{2020}\natexlab{}.
\newblock \showarticletitle{The trainer, the verifier, the imitator: Three ways
  in which human platform workers support artificial intelligence}.
\newblock \bibinfo{journal}{\emph{Big Data \& Society}} \bibinfo{volume}{7},
  \bibinfo{number}{1} (\bibinfo{year}{2020}),
  \bibinfo{pages}{2053951720919776}.
\newblock


\bibitem[Turner et~al\mbox{.}(2021)]%
        {turner2021human}
\bibfield{author}{\bibinfo{person}{Christopher~J Turner},
  \bibinfo{person}{Ruidong Ma}, \bibinfo{person}{Jingyu Chen}, {and}
  \bibinfo{person}{John Oyekan}.} \bibinfo{year}{2021}\natexlab{}.
\newblock \showarticletitle{Human in the Loop: Industry 4.0 technologies and
  scenarios for worker mediation of automated manufacturing}.
\newblock \bibinfo{journal}{\emph{IEEE access}}  \bibinfo{volume}{9}
  (\bibinfo{year}{2021}), \bibinfo{pages}{103950--103966}.
\newblock


\bibitem[Villemeur(1992)]%
        {villemeur1992assessment}
\bibfield{author}{\bibinfo{person}{Alain Villemeur}.}
  \bibinfo{year}{1992}\natexlab{}.
\newblock \bibinfo{title}{Assessment, hardware, software and human factors,
  volume 2 of Reliability, availability, maintainability and safety
  assessment}.
\newblock
\newblock


\bibitem[Walch et~al\mbox{.}(2016)]%
        {Walch2016TowardsCD}
\bibfield{author}{\bibinfo{person}{Marcel Walch}, \bibinfo{person}{Tobias
  Sieber}, \bibinfo{person}{Philipp Hock}, \bibinfo{person}{Martin R.~K.
  Baumann}, {and} \bibinfo{person}{Michael Weber}.}
  \bibinfo{year}{2016}\natexlab{}.
\newblock \showarticletitle{Towards Cooperative Driving: Involving the Driver
  in an Autonomous Vehicle's Decision Making}.
\newblock \bibinfo{journal}{\emph{Proceedings of the 8th International
  Conference on Automotive User Interfaces and Interactive Vehicular
  Applications}} (\bibinfo{year}{2016}).
\newblock


\bibitem[Wang et~al\mbox{.}(2022)]%
        {wang2022whose}
\bibfield{author}{\bibinfo{person}{Ding Wang}, \bibinfo{person}{Shantanu
  Prabhat}, {and} \bibinfo{person}{Nithya Sambasivan}.}
  \bibinfo{year}{2022}\natexlab{}.
\newblock \showarticletitle{Whose AI Dream? In search of the aspiration in data
  annotation.}. In \bibinfo{booktitle}{\emph{CHI Conference on Human Factors in
  Computing Systems}}. \bibinfo{pages}{1--16}.
\newblock


\bibitem[Wang et~al\mbox{.}(2020)]%
        {wang2020safety}
\bibfield{author}{\bibinfo{person}{Jun Wang}, \bibinfo{person}{Li Zhang},
  \bibinfo{person}{Yanjun Huang}, {and} \bibinfo{person}{Jian Zhao}.}
  \bibinfo{year}{2020}\natexlab{}.
\newblock \showarticletitle{Safety of autonomous vehicles}.
\newblock \bibinfo{journal}{\emph{Journal of advanced transportation}}
  \bibinfo{volume}{2020} (\bibinfo{year}{2020}).
\newblock


\bibitem[Wiegand et~al\mbox{.}(2020)]%
        {wiegand2020joy}
\bibfield{author}{\bibinfo{person}{Gesa Wiegand}, \bibinfo{person}{Kai
  Holl{\"a}nder}, \bibinfo{person}{Katharina Rupp}, {and}
  \bibinfo{person}{Heinrich Hussmann}.} \bibinfo{year}{2020}\natexlab{}.
\newblock \showarticletitle{The Joy of Collaborating with Highly Automated
  Vehicles}. In \bibinfo{booktitle}{\emph{12th International Conference on
  Automotive User Interfaces and Interactive Vehicular Applications}}.
  \bibinfo{pages}{223--232}.
\newblock


\bibitem[Wiegand et~al\mbox{.}(2019)]%
        {Wiegand2019ID}
\bibfield{author}{\bibinfo{person}{Gesa Wiegand}, \bibinfo{person}{Matthias
  Schmidmaier}, \bibinfo{person}{Thomas Weber}, \bibinfo{person}{Yuanting Liu},
  {and} \bibinfo{person}{Heinrich Hussmann}.} \bibinfo{year}{2019}\natexlab{}.
\newblock \showarticletitle{I Drive - You Trust: Explaining Driving Behavior Of
  Autonomous Cars}.
\newblock \bibinfo{journal}{\emph{Extended Abstracts of the 2019 CHI Conference
  on Human Factors in Computing Systems}} (\bibinfo{year}{2019}).
\newblock


\bibitem[Woods et~al\mbox{.}(1994)]%
        {woods1994behind}
\bibfield{author}{\bibinfo{person}{David~D Woods}, \bibinfo{person}{Leila~J
  Johannesen}, \bibinfo{person}{Richard~I Cook}, {and}
  \bibinfo{person}{Nadine~B Sarter}.} \bibinfo{year}{1994}\natexlab{}.
\newblock \bibinfo{booktitle}{\emph{Behind human error: Cognitive systems,
  computers and hindsight}}.
\newblock \bibinfo{type}{{T}echnical {R}eport}. \bibinfo{institution}{Dayton
  Univ Research Inst (Urdi) OH}.
\newblock


\bibitem[Xing et~al\mbox{.}(2021)]%
        {Xing2021TowardHC}
\bibfield{author}{\bibinfo{person}{Yang Xing}, \bibinfo{person}{Chen Lv},
  \bibinfo{person}{Dongpu Cao}, {and} \bibinfo{person}{Peng Hang}.}
  \bibinfo{year}{2021}\natexlab{}.
\newblock \showarticletitle{Toward human-vehicle collaboration: Review and
  perspectives on human-centered collaborative automated driving}.
\newblock \bibinfo{journal}{\emph{Transportation Research Part C: Emerging
  Technologies}} (\bibinfo{year}{2021}).
\newblock


\bibitem[Xu(2020)]%
        {xu2020automation}
\bibfield{author}{\bibinfo{person}{Wei Xu}.} \bibinfo{year}{2020}\natexlab{}.
\newblock \showarticletitle{From automation to autonomy and autonomous
  vehicles: Challenges and opportunities for human-computer interaction}.
\newblock \bibinfo{journal}{\emph{Interactions}} \bibinfo{volume}{28},
  \bibinfo{number}{1} (\bibinfo{year}{2020}), \bibinfo{pages}{48--53}.
\newblock


\bibitem[Xu et~al\mbox{.}(2023)]%
        {xu2023transitioning}
\bibfield{author}{\bibinfo{person}{Wei Xu}, \bibinfo{person}{Marvin~J Dainoff},
  \bibinfo{person}{Liezhong Ge}, {and} \bibinfo{person}{Zaifeng Gao}.}
  \bibinfo{year}{2023}\natexlab{}.
\newblock \showarticletitle{Transitioning to human interaction with AI systems:
  New challenges and opportunities for HCI professionals to enable
  human-centered AI}.
\newblock \bibinfo{journal}{\emph{International Journal of Human--Computer
  Interaction}} \bibinfo{volume}{39}, \bibinfo{number}{3}
  (\bibinfo{year}{2023}), \bibinfo{pages}{494--518}.
\newblock


\bibitem[Yang et~al\mbox{.}(2018c)]%
        {yang2018first}
\bibfield{author}{\bibinfo{person}{Shiyan Yang}, \bibinfo{person}{Steven~E
  Shladover}, \bibinfo{person}{Xiao-Yun Lu}, \bibinfo{person}{John Spring},
  \bibinfo{person}{David Nelson}, {and} \bibinfo{person}{Hani Ramezani}.}
  \bibinfo{year}{2018}\natexlab{c}.
\newblock \showarticletitle{A first investigation of truck drivers’
  on-the-road experience using cooperative adaptive cruise control}.
\newblock  (\bibinfo{year}{2018}).
\newblock


\bibitem[Yang et~al\mbox{.}(2018a)]%
        {Yang2018AnHC}
\bibfield{author}{\bibinfo{person}{Yucheng Yang}, \bibinfo{person}{Burak
  Karakaya}, \bibinfo{person}{Giancarlo~Caccia Dominioni},
  \bibinfo{person}{Kyosuke Kawabe}, {and} \bibinfo{person}{Klaus Bengler}.}
  \bibinfo{year}{2018}\natexlab{a}.
\newblock \showarticletitle{An HMI Concept to Improve Driver's Visual Behavior
  and Situation Awareness in Automated Vehicle}.
\newblock \bibinfo{journal}{\emph{2018 21st International Conference on
  Intelligent Transportation Systems (ITSC)}} (\bibinfo{year}{2018}),
  \bibinfo{pages}{650--655}.
\newblock


\bibitem[Yang et~al\mbox{.}(2018b)]%
        {yang2018hmi}
\bibfield{author}{\bibinfo{person}{Yucheng Yang}, \bibinfo{person}{Burak
  Karakaya}, \bibinfo{person}{Giancarlo~Caccia Dominioni},
  \bibinfo{person}{Kyosuke Kawabe}, {and} \bibinfo{person}{Klaus Bengler}.}
  \bibinfo{year}{2018}\natexlab{b}.
\newblock \showarticletitle{An hmi concept to improve driver's visual behavior
  and situation awareness in automated vehicle}. In
  \bibinfo{booktitle}{\emph{2018 21st International Conference on Intelligent
  Transportation Systems (ITSC)}}. IEEE, \bibinfo{pages}{650--655}.
\newblock


\bibitem[Zhang et~al\mbox{.}(2021)]%
        {zhang2021human}
\bibfield{author}{\bibinfo{person}{Jiehuang Zhang}, \bibinfo{person}{Ying Shu},
  {and} \bibinfo{person}{Han Yu}.} \bibinfo{year}{2021}\natexlab{}.
\newblock \showarticletitle{Human-Machine Interaction for Autonomous Vehicles:
  A Review}. In \bibinfo{booktitle}{\emph{International Conference on
  Human-Computer Interaction}}. Springer, \bibinfo{pages}{190--201}.
\newblock


\end{thebibliography}

\appendix

\end{document}